\documentclass[twocolumn]{aastex631}

\usepackage{xcolor,tikz}
\usepackage{newtxtext,newtxmath,soul}
\usepackage{multirow}
\usepackage[T1]{fontenc}
\usepackage{gensymb}
\usepackage{tablefootnote}
\usepackage{makecell}
\usepackage{comment}
\usepackage{amsmath}
\usepackage{graphicx}

\begin{document}

\title{Magnetic White Dwarfs in the SDSS 100 pc Sample: Further Evidence of Two Formation Channels}
\shorttitle{Magnetic WDs}

\author[0000-0001-7143-0890]{Adam Moss}
\affiliation{Homer L. Dodge Department of Physics and Astronomy, University of Oklahoma, 440 W. Brooks St., Norman, OK 73019, USA}
\author[0000-0001-6098-2235]{Mukremin Kilic}
\affiliation{Homer L. Dodge Department of Physics and Astronomy, University of Oklahoma, 440 W. Brooks St., Norman, OK 73019, USA}
\author[0000-0003-2368-345X]{Pierre Bergeron}
\affiliation{Département de Physique, Université de Montréal, C.P. 6128, Succ. Centre-Ville, Montréal, Québec H3C 3J7, Canada}
\author[0000-0002-4462-2341]{Warren R. Brown}
\affiliation{Center for Astrophysics | Harvard \& Smithsonian,
60 Garden Street, Cambridge, MA 02138, USA}
\author[0009-0009-9105-7865]{Gracyn Jewett}
\affiliation{Homer L. Dodge Department of Physics and Astronomy, University of Oklahoma, 440 W. Brooks St., Norman, OK 73019, USA}
\author[0000-0001-7077-3664]{Marcel A. Agüeros}
\affiliation{Department of Astronomy, Columbia University, 550 West 120th Street, New York, NY 10027, USA}
\author[0000-0002-3524-190X]{Maria Camisassa}
\affiliation{Departament de Física, Universitat Politècnica de Catalunya, c/Esteve Terrades 5, 08860 Castelldefels, Spain}
\author[0000-0002-5380-0816]{Anthony Burrow}
\affiliation{Homer L. Dodge Department of Physics and Astronomy, University of Oklahoma, 440 W. Brooks St., Norman, OK 73019, USA}
\shortauthors{Moss et al.}
\email{adam.g.moss-1@ou.edu}

\begin{abstract}

We conduct a model atmosphere analysis on all magnetic white dwarfs in the SDSS 100 pc sample. We have 163 magnetic targets in this sample, 87 of which are new discoveries, making this the largest volume-limited survey of magnetic white dwarfs to date. We discuss the distribution of multiple parameters, including mass, cooling age, and field strength. We find strong evidence of two populations of magnetic white dwarfs that form through separate mechanisms based on a cluster analysis of these parameters. The young, high mass objects typically have high field strengths which indicate a merger origin, while old, average mass objects have weaker fields that likely originated through a crystallization-induced dynamo or previous evolution stages. When comparing young and old objects, two-sample Kolmogorov-Smirnov tests yield statistically significant differences between the field strengths and masses of the magnetic targets. We use a Gaussian mixture model to identify where these populations lie in parameter space, and we find two groups centered at distinct cooling ages, masses, and field strengths: 2.9 Gyr, 0.71 $M_{\odot}$, 3.7 MG and 1.8 Gyr, 0.96 $M_{\odot}$, 84 MG respectively. Our results further support the dual formation channel previously reported in the literature. The occurrence of magnetism strongly correlates with the onset of crystallization. However, given the breakout times required for a crystallization dynamo, we find that many of our older, average mass objects can be better explained with a core-convective dynamo that forms on the main-sequence. 

\end{abstract}


\section{Introduction}
The origin of magnetic fields in white dwarfs (WDs) remains an open question over 50 years since the first discovery of a surface magnetic field in a WD \citep{Kemp70,Angel71,Landstreet71,Angel72}. Multiple channels have been proposed to explain the incidence of magnetism across a wide range of field strengths, effective temperatures, and WD masses, with no single mechanism able to successfully explain all magnetic white dwarfs (MWDs, see \citealt{Ferrario15,Ferrario20}, for a review). Thus the source of magnetism in many objects is unclear.

The three primary explanations for the generation of a magnetic field are a fossil origin, crystallization, and binary interactions. In the fossil case, the WD retains a magnetic field from prior evolution stages \citep{Tout04,Cantiello16}. While the incidence of highly magnetic Ap/Bp stars is not high enough to account for all MWDs \citep{Wick05}, recent work has shown that the breakout time for a magnetic field generated by a convective core on the main-sequence indeed aligns with the WD phase \citep{Camisassa24}. For a crystallization-induced dynamo, convective flows generated by the oxygen enrichment of the solidifying core could sustain a magnetic field \citep{Isern17,Schreiber21,Ginzburg22}. In contrast to the main-sequence dynamo, however, \citet{Blatman24} showed that it can take $\sim$1$-$7~Gyrs before the field emerges at the surface, which is longer than many of the MWD cooling ages determined in volume-limited samples \citep{Bagnulo21,Bagnulo22}. Lastly, close binary evolution such as a double WD merger \citep{Garcia12} or post-common-envelope evolution \citep{Tout08,Briggs18} is a likely source of the most massive MWDs and those with the strongest fields, especially since MWDs tend to have shorter rotation periods than their non-magnetic counterparts \citep{Hernandez24,Oliveira24}.

Adding to the complexity are the detection methods for discovering MWDs, which have inherent biases that prevent a complete understanding of the population. The detection of Zeeman-splitting in optical absorption lines is perhaps the simplest way, with the bulk of discoveries coming from large spectroscopic surveys such as the Sloan Digital Sky Survey (SDSS) \citep{Kepler13}. However, low-resolution spectroscopy will fail to detect fields weaker than $\sim$1 MG \citep{Kepler13}. To detect weaker fields, high-resolution spectroscopy or spectropolarimetry are required (see \citealt{Landstreet15} for a summary of the diagnostic content of spectropolarimetry). In addition, the use of Zeeman-splitting becomes impossible as absorption features in the spectra disappear once the WD reaches cool enough effective temperatures. Spectropolarimetry or broadband polarimetry are thus required to detect magnetic fields in WDs with featureless spectra (DC WDs). 

Thanks to unbiased volume-limited surveys, we are starting to obtain a more holistic view of the population and improve our understanding of the origins of magnetic fields in WDs. \citet{Bagnulo21} conducted a spectropolarimetric survey and found 33 MWDs out of 149 total within the 20 pc volume. The authors made several important findings, namely the rarity of young MWDs, the increase in frequency of magnetism with cooling age, and the constant distribution of field strength with cooling age. Given the lack of young MWDs in this sample, \citet{Bagnulo22} looked at all WDs younger than 0.6 Gyr and within 40 pc of the Sun. They found clear evidence of two distinct populations of MWDs. The first group consists of the massive ($\geq1.1 M_{\odot}$) and young ($\tau < 0.6$ Gyr) WDs which are frequently magnetic, have the strongest fields among magnetic objects, and whose fields are detectable at the surface quite early in the cooling phase. The second group are those below 0.75$M_{\odot}$ in which the fields do not emerge at the surface until much later in the cooling sequence. The authors also found that the incidence and strength of fields in these average mass objects grow over time. These two populations likely have separate contributing formation channels: the former group likely receives significant contribution from binary interactions, while those in the latter group likely come exclusively from single-star evolution. These findings had similarly been predicted by \citet{Valyavin99}.

Here we investigate the origin of MWDs using the 100 pc sample in the SDSS footprint. This is an extension of \citet{Kilic25}, who targeted WDs hotter than 5000 K with optical spectroscopy, and achieved a 86\% spectroscopic completeness for targets hotter than 5000 K and 91\% for those hotter than 6000 K. We discuss our observations and sample selection in Section 2, followed by our model atmosphere analysis in Section 3. In Section 4 we discuss the results with respect to various physical features such as magnetic field strength, photometric variability, cooling age, etc. In Section 5 we show statistical evidence for two populations of MWDs in our sample, and then discuss our findings with regards to the unsettled questions in WD magnetism in Section 6. We then conclude in Section 7. 

\section{Sample Selection and Observations}

As described in \citet{Kilic25}, there are 3146 WDs with spectral classifications out of 4214 in the entire SDSS 100 pc sample. These 3146 targets consist of 2306 WDs with spectral classification from the literature, and 840 obtained from follow-up optical spectroscopy. The spectroscopic follow-up is 86\% complete for targets with $T_{\rm eff} \geq 5000$ K and 91\% for targets with $T_{\rm eff} \geq 6000$ K. The follow-up observations were conducted using the 1.5m Fred Lawrence Whipple Observatory (FLWO), MDM Hiltner 2.4m, Apache Point Observatory (APO) 3.5m, 6.5m MMT, 6.5m Magellan, and 8m Gemini South telescopes between 2020 October and 2024 May. MDM observations were obtained as part of the OSMOS queue, and Gemini data were taken as part of the queue programs GS-2022B-Q- 304, GS-2023A-Q-227, and GS-2023A-Q-327.

We primarily identify magnetic targets based on Zeeman-splitting of optical absorption lines. Since a high magnetic field strength ($\sim$100 MG) can smear out absorption lines, we can also detect fields if a target has a featureless spectrum and if it is at a sufficiently high effective temperature such that we would expect to see H or He lines. We do not conduct spectropolarimetry in our observations, thus we are biased against objects with very weak fields ($B < 1$ MG) and DCs at lower effective temperatures that could harbor high field strengths ($B > 100$ MG).

From this sample, we identify 163 total MWDs, 149 of which have hydrogen-dominated atmospheres (91\%). This sample is thus the largest volume-limited sample of MWDs to date. The magnetic fraction however is notably lower (5.2\%) than both \citet{Bagnulo21} and \citet{Bagnulo22}, who obtained a magnetic fraction of 22\% for the local 20 pc volume. This is not surprising given the limitations of using optical spectroscopy only to detect magnetic fields. If we only consider objects with field strengths 1 $\leq B \leq$ 100 MG from \citet{Bagnulo22}, the magnetic fraction drops to 5.9\%. Hence our magnetic fraction is consistent given our biases against very weak and very strong fields. Note that there are two objects with Zeeman-split Balmer lines in emission (DAHe) in the 100 pc sample. Since the origin of these targets is unclear \citep{Manser23}, we do not include them in our analysis. 

\begin{figure}[!ht]
    \centering
    \includegraphics[width=3.75in, clip=true, trim=0.2in 0in 0in 0in]{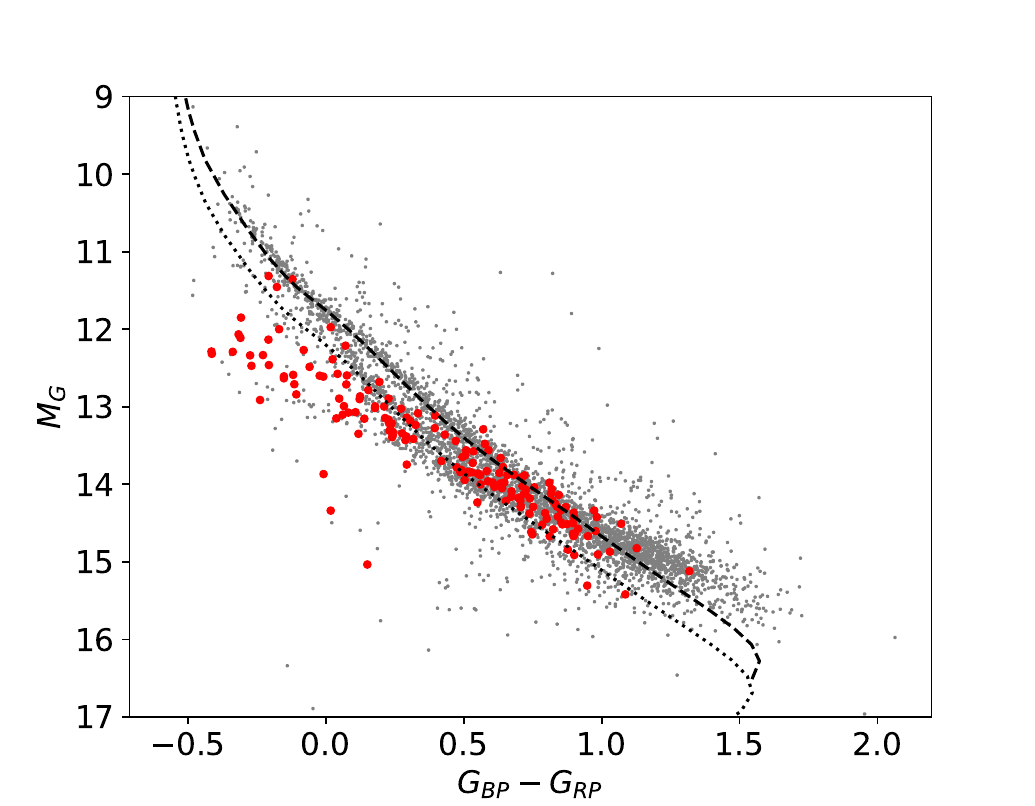}
    \caption{Gaia color-magnitude diagram of the 100 pc sample in the SDSS footprint (gray points) with the MWDs marked by red points. The dashed and dotted lines are the cooling tracks for a $0.6 M_{\odot}$ and $0.8 M_{\odot}$ WD respectively with pure H atmospheres, which go down to $T_{\rm eff} = 3000$ K in this figure.}
    \label{fig1} 
\end{figure}

Figure \ref{fig1} shows the Gaia color-magnitude diagram of the 100 pc sample with the magnetic targets highlighted and the evolutionary sequences for a $0.6 M_{\odot}$ and $0.8 M_{\odot}$ WD with pure H atmospheres \citep{Tremblay11,Blouin19,Bedard20}. We immediately see an above-average mass tendency in hotter MWDs, while the cooler ones are more concentrated among average mass WDs. This is in line with the results from \citet{Bagnulo22}, who found that the mass of MWDs with cooling age $\geq 2$~Gyrs is similar to the mass of normal WDs. We also see a lack of MWDs redder than $G_{BP}-G_{RP}\approx 1.0$ in our sample. This corresponds to $T_{\rm eff}=5000$ K, below which H lines disappear. We cannot detect magnetism in WDs redder than this limit through optical spectroscopy alone, so we are biased against the oldest and coolest MWDs. Thus our observations provide a robust lower limit on the number of MWDs in the sample.  

\section{Model Atmosphere Analysis}

Of the 149 MWDs with hydrogen-dominated atmospheres, 81 of these are new discoveries while 2 were found by \citet{Berdyugin24} with no field estimate provided. For these 83 targets, we use offset dipole models to obtain fits to our spectra. We start by using the photometric technique described by \citet{Bergeron19} to constrain the effective temperature and solid angle, and thus the radius and mass. We obtain these parameters for all 163 MWDs in our sample. For the 83 new objects, we construct a grid of magnetic spectra to determine the viewing angle $i$ ($i=0\degree$ means viewing the WD ``pole-on''), magnetic field strength $B_{d}$, and dipole offset $a_{z}$ (see Figure 1 of \citealt{Achilleos89} for the 3D model of this geometry and Figure 4 of \citealt{Bergeron92} for example synthetic spectra). We do not provide model fits for the 6 newly found non-DAs (2 DQHs and 4 DZHs).

The total line opacity is calculated as the sum of the individual Stark-broadened Zeeman components. The line displacements and oscillator strengths of the Zeeman components of H$\alpha$ through H$\delta$ are kindly provided by S. Jordan (see also \citealt{Hardy23a}). The total line opacity is normalized to that resulting from the zero-field solution. The specific intensities at the surface, $I(\nu,\mu,\tau_{\nu}=0)$, are obtained by solving the radiative transfer equation for various field strengths and values of $\mu$ ($\mu= \cos \theta$, where $\theta$ is the angle between the propagation of light and the normal to the surface of the star). Further details of these magnetic models are provided in \citet{Moss24}.

\begin{figure}[!ht]
    \centering
    \includegraphics[width=3.7in, clip=true, trim=1in 3in 0.1in 3.4in]{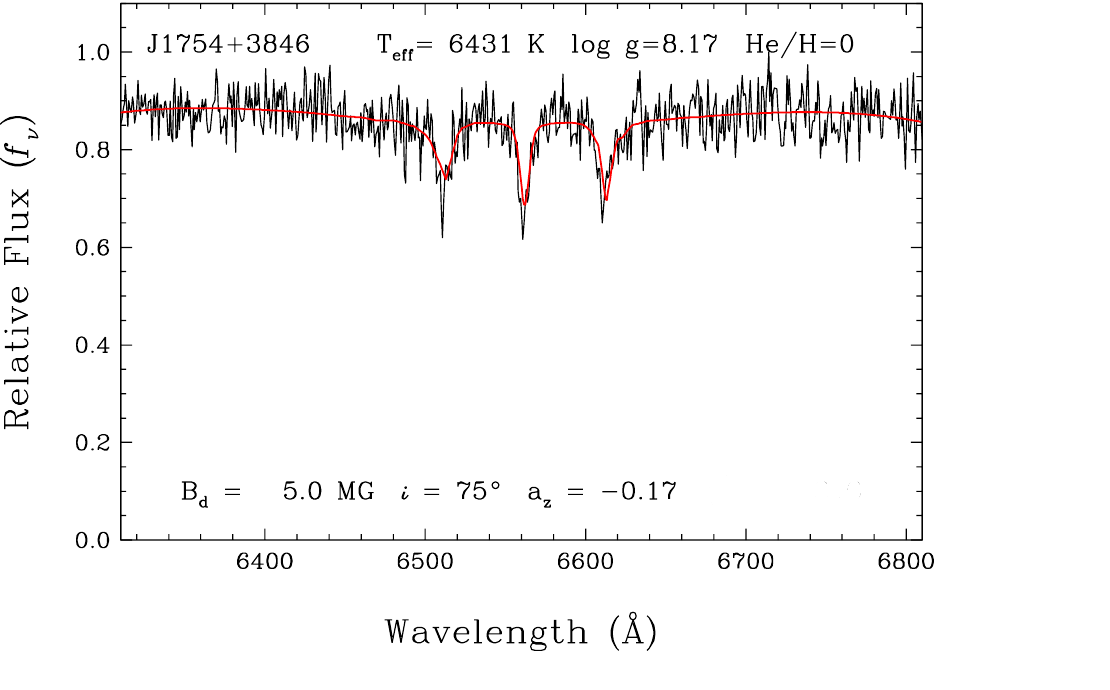}
    \caption{Offset dipole fit to the H$\alpha$ line of a DAH, J1754+3846. We obtain an excellent fit with a magnetic field strength of 5.0 MG.}
    \label{fig2} 
\end{figure}

We typically obtain excellent fits for field strengths 2~MG~$\lesssim~B_d~\lesssim$~100~MG. Above 100 MG, the absorption lines tend to smear out or become distorted, which makes determining the specific geometry difficult. Figure \ref{fig2} shows an example fit to the H$\alpha$ line of a DAH with an average magnetic field strength. We obtain a field strength of 5.0 MG for this target, which yields an excellent fit to the central H$\alpha$ line and the Zeeman-split components. Thus we are confident that the field strength and geometry for this object are well constrained.  

\begin{figure}[!ht]
    \centering
    \includegraphics[width=3.7in, clip=true, trim=1in 3in 0.1in 3.4in]{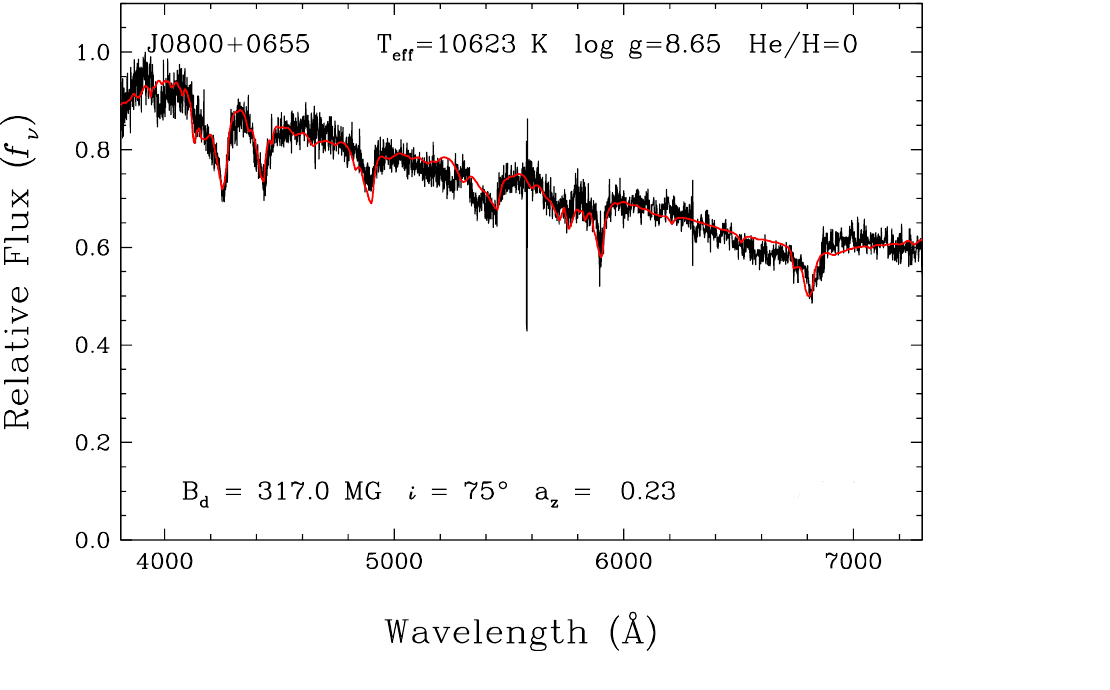}
    \includegraphics[width=3.7in, clip=true, trim=1in 3in 0.1in 3.4in]{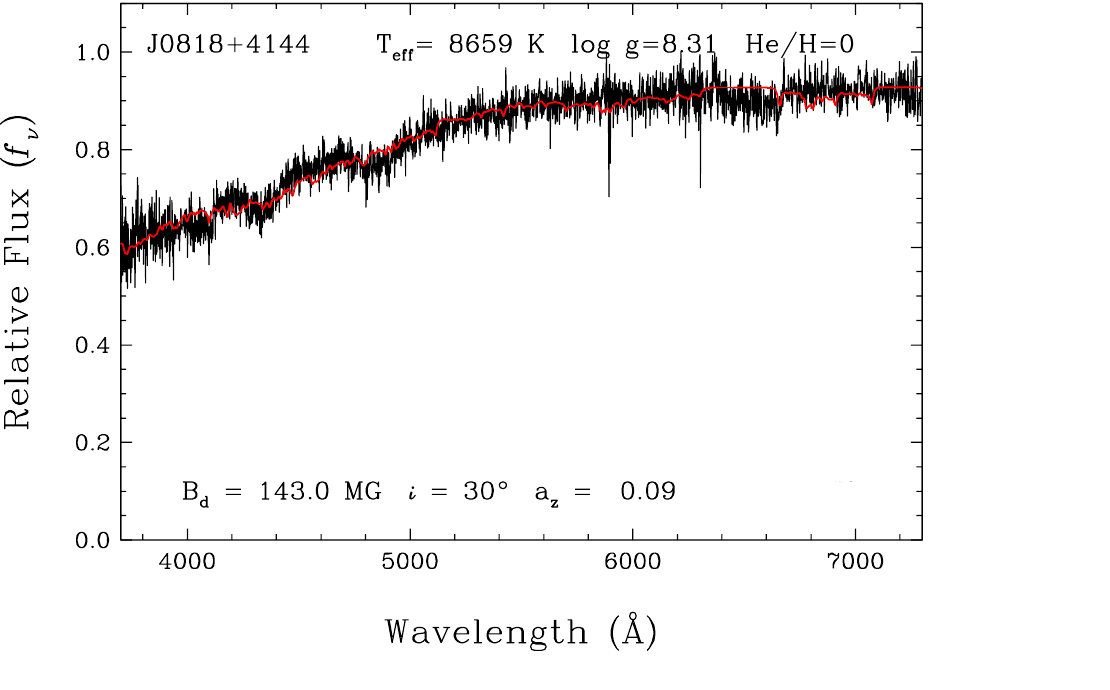}
    \caption{Fits to two targets that possess some of the highest field strengths in our sample. For J0800+0655 (top), absorption lines are present but significantly shifted from their zero-field positions. In contrast, the lines appear mostly smeared out for J0818+4144 (bottom).}
    \label{fig3} 
\end{figure}

Figure \ref{fig3} shows the fits to two targets where we obtain field strengths above 100 MG. For J0800+0655, the absorption features are shifted but still prominent. We are thus able to obtain an excellent fit with a field strength of 317 MG, and we consider this field strength accurately constrained. In contrast, the absorption features of J0818+4144 are more smeared out, likely due to the high strength and geometry of the field. While the field strength is likely above 100 MG, we consider this value questionable as we are not confident that 143 MG is the exact field strength of this object based on the quality of our fit.  

\begin{figure}[!ht]
    \centering
    \includegraphics[width=3.3in, clip=true, trim=0.4in 3in 0in 3.3in]{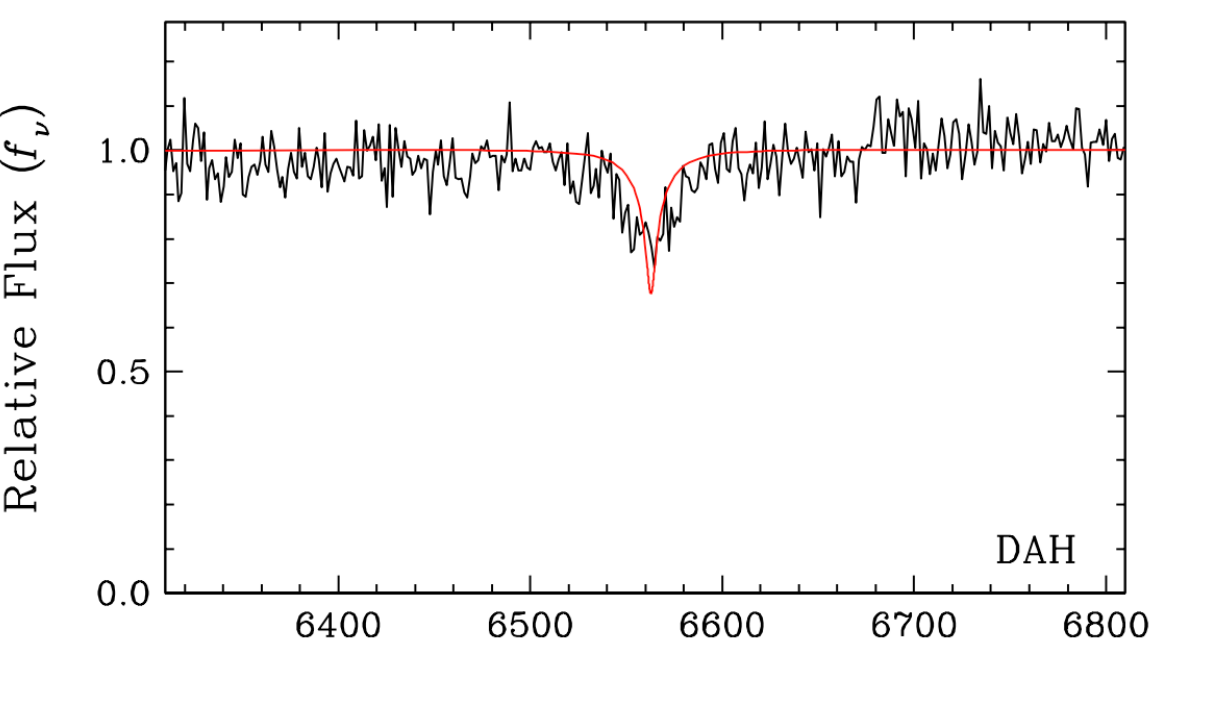}
    \includegraphics[width=3.7in, clip=true, trim=1in 3in 0.1in 3.4in]{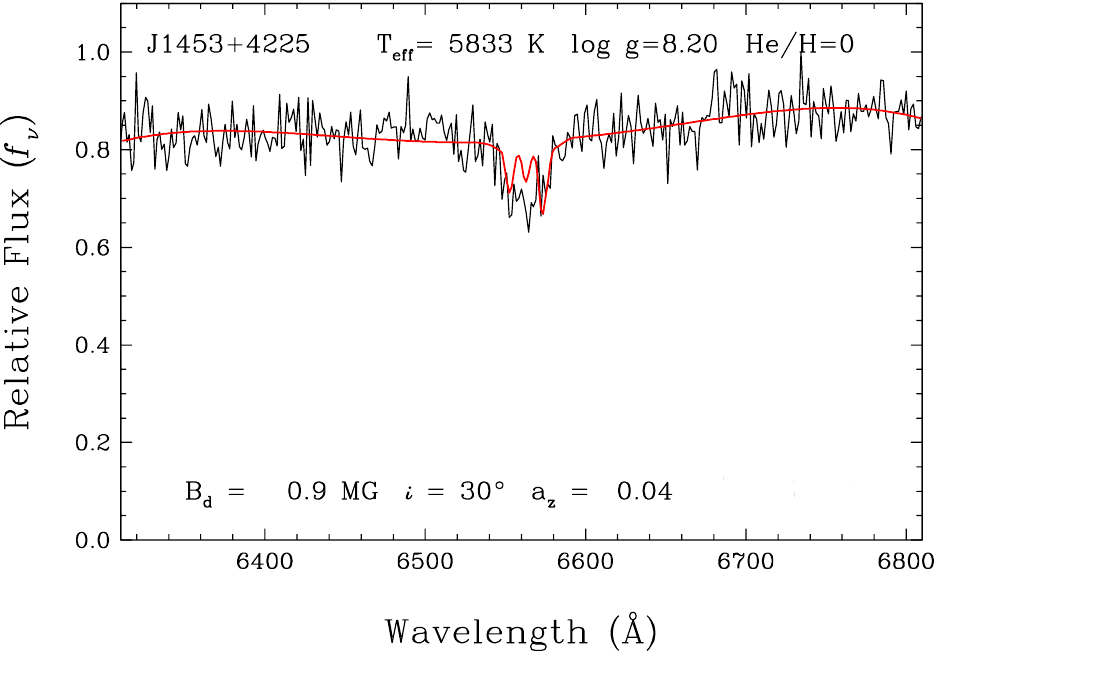}
    \caption{An example fit to a target with a low field strength ($< 1$ MG). While Zeeman-splitting is not obvious, the shape of the line profile compared to that of a typical DA suggests the presence of a magnetic field.}
    \label{fig4} 
\end{figure}

Below a few MG, the Zeeman-split components are not shifted far from the central H$\alpha$ line, making them hard to distinguish when fitting the spectra. However, we can infer the presence of a field based on the non-magnetic fit to H$\alpha$. Figure \ref{fig4} shows an example non-magnetic and magnetic fit to an object with a field strength on the order of 1 MG. Here we do not see obvious Zeeman-splitting, but the shape of the H$\alpha$ line is clearly distorted from what we would expect from a DA. Hence we classify this target as magnetic, but we mark the field strength as questionable. 

\begin{figure}[!ht]
    \centering
    \includegraphics[width=3.7in, clip=true, trim=0.5in 3in 0.1in 2.9in]{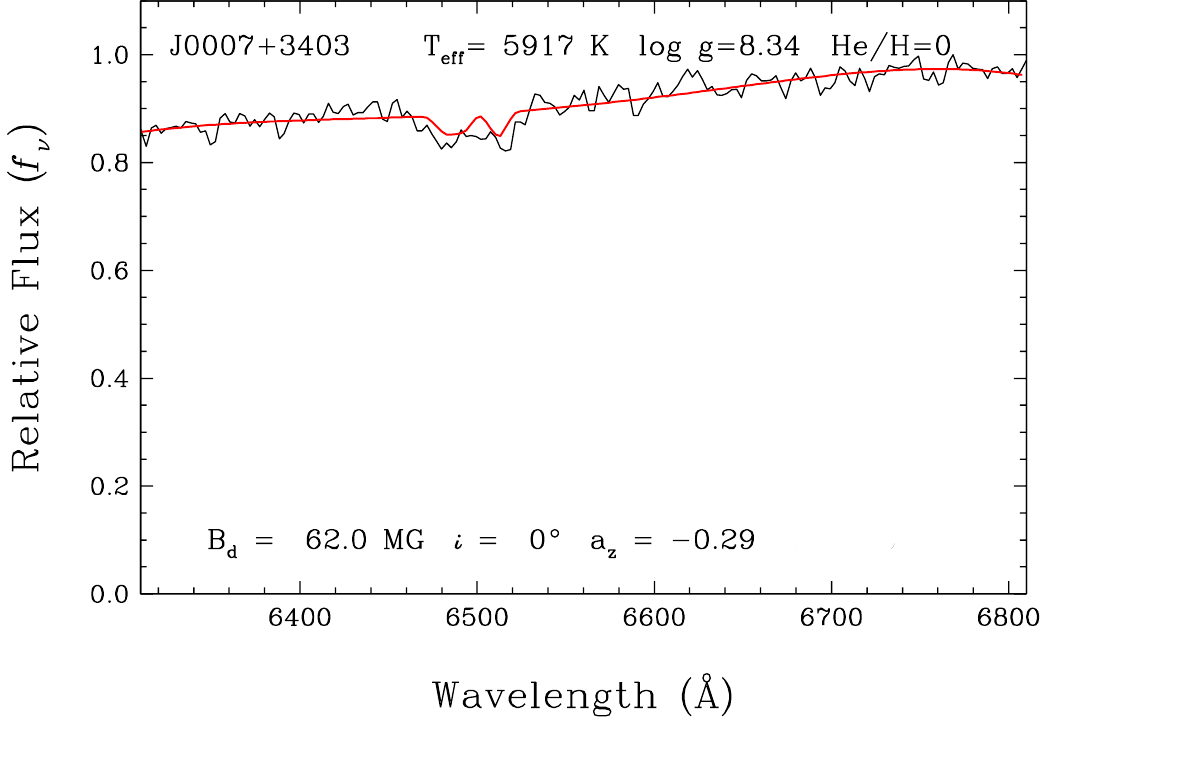}
    \caption{Offset dipole fit to a DAP in our sample, which was recently classified as a DCP. The H$\alpha$ line is weak but noticeable.}
    \label{fig5} 
\end{figure}

We also identify a DAP in our sample, SDSSJ000728.91+340341.4. This target was previously classified as a DCP \citep{Berdyugin24}, however we detect a weak H$\alpha$ feature in the optical spectrum from \citet{Limoges15}. Hence we classify this object now as a DAP. We use the spectrum obtained by \citet{Limoges15} to constrain the magnetic field strength. We obtain a magnetic field strength of $B_d = 62$ MG in our fit, shown in Figure \ref{fig5}.

Of the 81 new discoveries, there are 13 in which the presence of a magnetic field is likely but unclear, which we classify as ``DAH:''. There are also 9 targets where we are unsure of the atmospheric composition, which we label as ``DAH?''. We consider the field strengths in these spectral types preliminary. We also have 14 magnetic non-DAs in our sample, including 1 DBH, 1 DBAH, 1 DCP, 4 DQH, and 7 DZH WDs.  4 of the DZHs are new identifications, and their spectra are shown in \citet{Kilic25}. The DCP, SDSSJ074842.48+112502.0, was recently identified as magnetic  by \citet{Berdyugin24} using broadband polarimetry. However, no magnetic field estimate was given in that study, so we provide a fit assuming a hydrogen-dominated atmosphere. Given the lack of strong absorption lines, we provide a tentative fit which yields a field strength of $B_d = 112$ MG. The exact field strength is likely much higher. The fits to the newly identified MWDs, as well as the two DAP and DCP WDs, can be found on Zenodo, which can be accessed via the DOI \url{https://doi.org/10.5281/zenodo.15163601}.

We have magnetic field strength estimates for 152 of the 163 objects in the sample. 83 (81 new DAHs, 1 DAP and 1 DCP) of these come from our own analysis, while the remaining are taken from the literature. We list the physical parameters for all 163 MWDs in Table \ref{tab3} in the Appendix. We inspect each of our fits and note the dipole parameters that we consider preliminary from our analysis with an ``X'' in the final column. We do not expect this to affect our results, which depend more on the order of magnitude of the field strengths rather than the specific values themselves. However, the fits in these cases are not sufficient enough for us to conclude that the parameters we obtain represent the exact geometry of the objects. We also do not make a distinction between different types of field measurements in our analysis. For example, we use the mean field modulus $\langle|B|\rangle$ obtained by \citet{Bagnulo24} for SDSSJ162157.77+043218.8, while our model atmosphere analysis and most of the estimates in the literature use the dipole field strength $B_d$. Similar to our preliminary fits, we do not expect this lack of distinction to impact our final results. Lastly, our theoretical framework follows that of \citet{Hardy23a}, who found relative uncertainties of 5\% in the field strengths, 20\% for the viewing angle, and 50\% for the dipole offset. These uncertainties were determined by varying the parameters until the resulting synthetic spectrum significantly diverged from the observed spectrum. Hence we do not report exact error estimates for these parameters in Table \ref{tab3}.

\begin{figure}[!ht]
    \centering
    \includegraphics[width=3.5in, clip=true, trim=0in 0in 0in 0in]{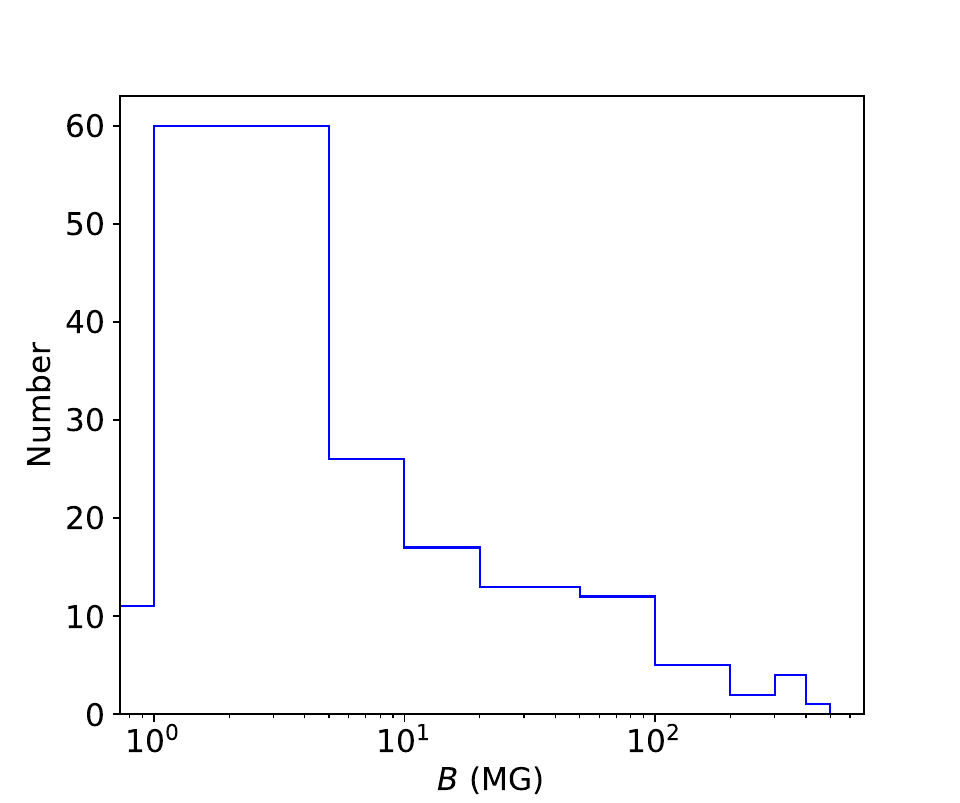}
    \caption{Distribution of field strengths in the SDSS 100 pc sample. The sample is dominated by intermediate fields between 1 and 10 MG, with a median field strength of 5.2 MG. The sharp decline below 1 MG is likely an observational bias in our sample which relies on low-resolution spectroscopy.} 
    \label{fig6}
\end{figure}

Figure \ref{fig6} shows the distribution of field strengths. Our sample is dominated by lower field objects, specifically those between 1 and 10 MG. There is a sharp decline in the number of objects below 1 MG, which is a bias in our sample. Since we primarily identify targets as magnetic based on Zeeman-splitting, and many of the objects do not have spectropolarimetry observations in the literature, we are biased against the weakest fields ($B < 1$ MG). A summary of the MWDs in our sample is presented in Table \ref{tab1}. 

\begin{deluxetable}{cc}
    \tabletypesize{\normalsize}
    \tablecolumns{2} \tablewidth{0pt}
    \centering
    \caption{\label{tab1} Summary of our MWD numbers for a given category.}
    \tablehead{\colhead{Category} & \colhead{Number}\\}
    \startdata
           Total  & 163\\
           Hydrogen-dominated Atmospheres & 149\\
           Non-DA & 14\\
           Field Strength Measurements & 152\\
           Newly Discovered Objects & 87\\
           New Field Measurements & 83\\
           Preliminary Fits & 47\\
           DAH: & 13\\
           DAH? & 9\\
    \enddata    
\end{deluxetable}

\section{Results}

\subsection{Effective Temperature}
\begin{figure*}[!ht]
    \centering
    \includegraphics[width=4.5in, angle=90, clip=true, trim=1.5in 1in 1.5in 1in]{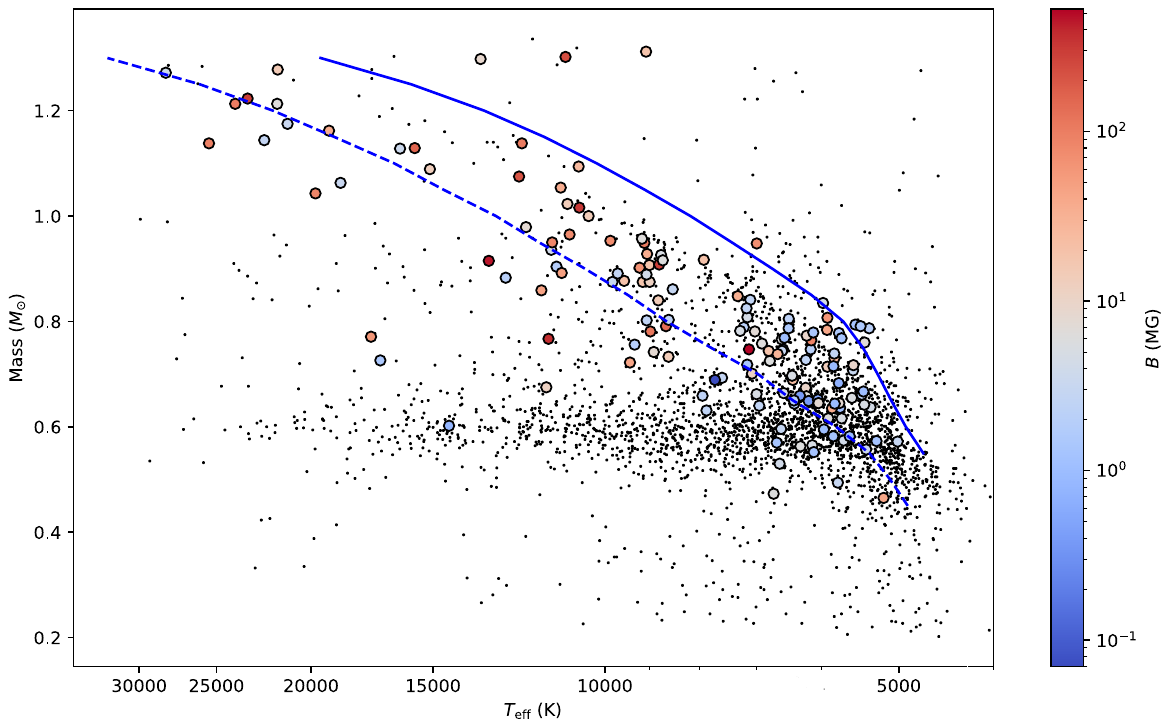}
    \caption{Stellar masses as a function of effective temperature for all spectroscopically confirmed WDs in the SDSS 100 pc sample. Objects that are classified as magnetic in the literature and in this work are colored by field strength. We have field strength measurements for 152 of the 163 MWDs in the sample. The blue tracks correspond to the onset of crystallization (dashed line) and 80\% core crystallization (solid line) assuming a carbon-oxygen core composition.}
    \label{fig7} 
\end{figure*}

Our goal is to analyze the trends in magnetism across multiple parameters in order to constrain the formation channel and evolution of magnetic fields in WDs. Figure \ref{fig7} shows the stellar masses as a function of effective temperature of the 100 pc sample. The MWDs are marked with large circles, and we indicate their field strength measurements by color. There appears to be a clear trend in the occurrence of magnetism and the onset of crystallization, with $\approx$2/3 of the magnetic objects having begun core crystallization. The exact start of crystallization depends on the oxygen abundance in the core. These evolutionary models assume equal amounts of C and O. However, a higher oxygen abundance \citep[e.g. $\sim$0.8,][]{Giammichele18,Giammichele22} would shift the onset to higher temperatures, which would result in a larger fraction of our magnetic sample undergoing crystallization. 

\citet{Amorim23} used offset dipole models to estimate the field strength of 804 DAHs found in the SDSS up to DR16 and determined that almost all objects with field strengths above 100 MG have begun core crystallization (see their Figure 13). However, the majority of their DAHs with field strengths weaker than 100 MG have not yet undergone crystallization in contrast with our sample. This is expected as the \citet{Amorim23} sample is magnitude-limited such that objects with low mass and high $T_{\rm eff}$, which begin crystallization later in the cooling sequence, are intrinsically brighter and thus over-represented.

\begin{figure*}[!ht]
    \centering
    \includegraphics[width=4.5in, angle=90, clip=true, trim=1.75in 1in 1.5in 1in]{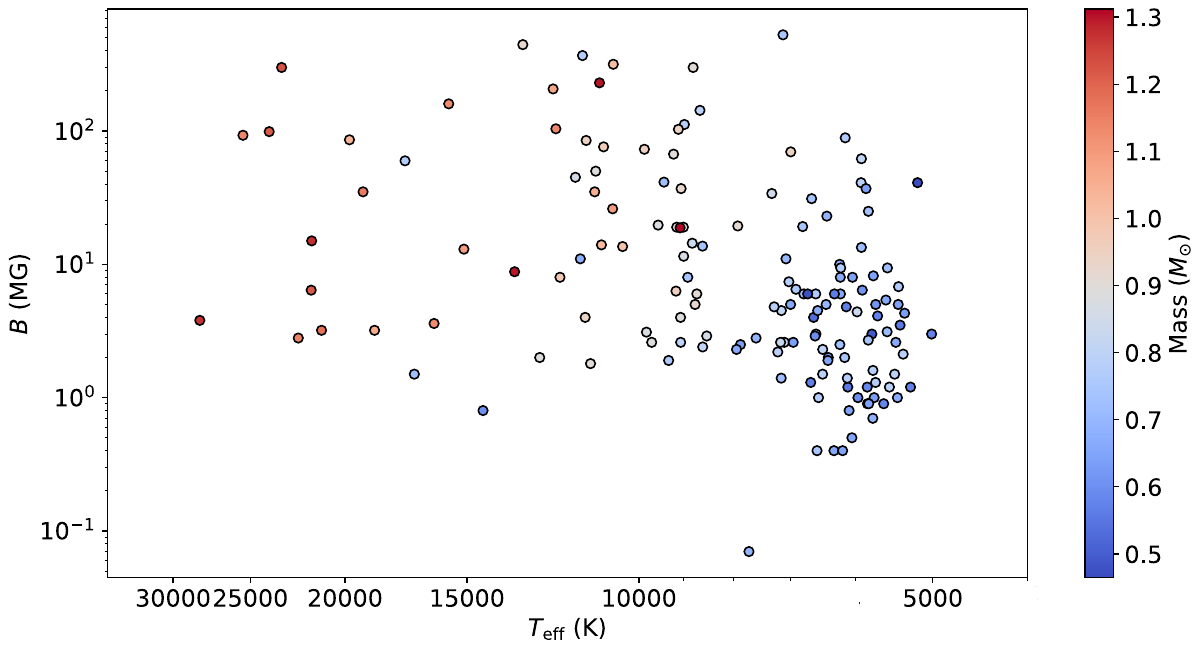}
    \caption{Field strength as a function of effective temperature, with mass indicated by color. There is a noticeable distinction between objects that are hotter/colder than $\sim$8000 K, with cooler objects having lower mass than hotter objects.} 
    \label{fig8}
\end{figure*}

We also see a loose trend in the field strengths, with weaker fields becoming prevalent at low mass and effective temperature. This becomes more apparent in Figure \ref{fig8} where we plot the magnetic field strength vs effective temperature and indicate mass by color. We can visually see two possible populations of MWDs: those cooler than $\sim$8000 K that tend to have field strengths weaker than $\sim$10 MG, and those at hotter temperatures that have a wide range of field strengths including some of the strongest in our sample. In addition, the magnetic targets at lower temperatures exclusively have lower masses. 

\subsection{Mass}

\begin{figure}[!ht]
    \centering
    \includegraphics[width=3.5in, clip=true, trim=0in 0in 0in 0in]{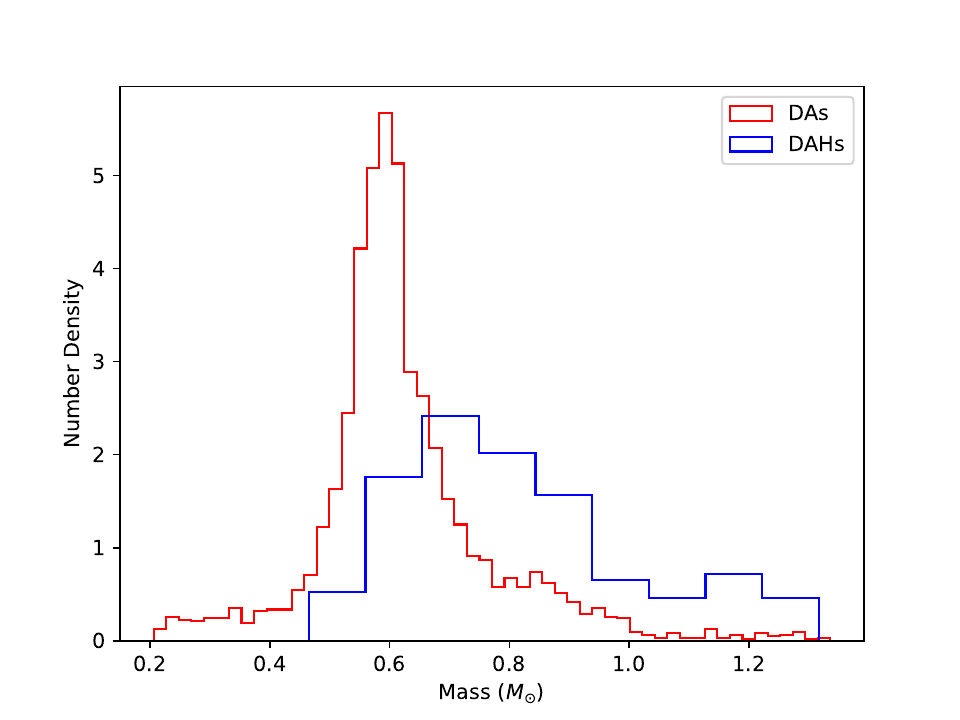}
    \caption{Mass distributions of the hydrogen-dominated atmosphere WDs in the 100 pc sample. Each histogram is normalized such that the area under the curve is equal to one. The non-magnetic distribution peaks at 0.59 $M_{\odot}$ with a broad shoulder, while the magnetic distribution peaks at 0.7 $M_{\odot}$ and has a high mass tail.} 
    \label{fig9}
\end{figure}

Figure \ref{fig9} shows the mass distribution of the non-magnetic and MWDs with hydrogen-dominated atmospheres. As shown in many previous studies, the magnetic objects have higher masses on average than the non-magnetic targets \citep{Liebert03,Kepler13,Ferrario15,Mccleery20,Amorim23,Hardy23a,Obrien24}. While the peak of the magnetic distribution occurs at 0.7 $M_{\odot}$, we see a high mass tail which consists of many likely merger remnants. \citet{Temmink20} performed a population synthesis study on the single WDs within 100 pc and found that 30$-$50\% of massive WDs originate from binary mergers, but only 10$-$30\% for WDs of all masses. \citet{Jewett24} analyzed all WDs greater than $0.9M_{\odot}$ in the Pan-STARRS footprint within 100 pc and found that the merger fraction ranges from 25\% to 78\% for different masses. Hence mergers are a more common source of these high mass MWDs compared to those at lower mass.

\begin{figure}[!ht]
    \centering
    \includegraphics[width=3.75in, clip=true, trim=0.2in 0in 0in 0in]{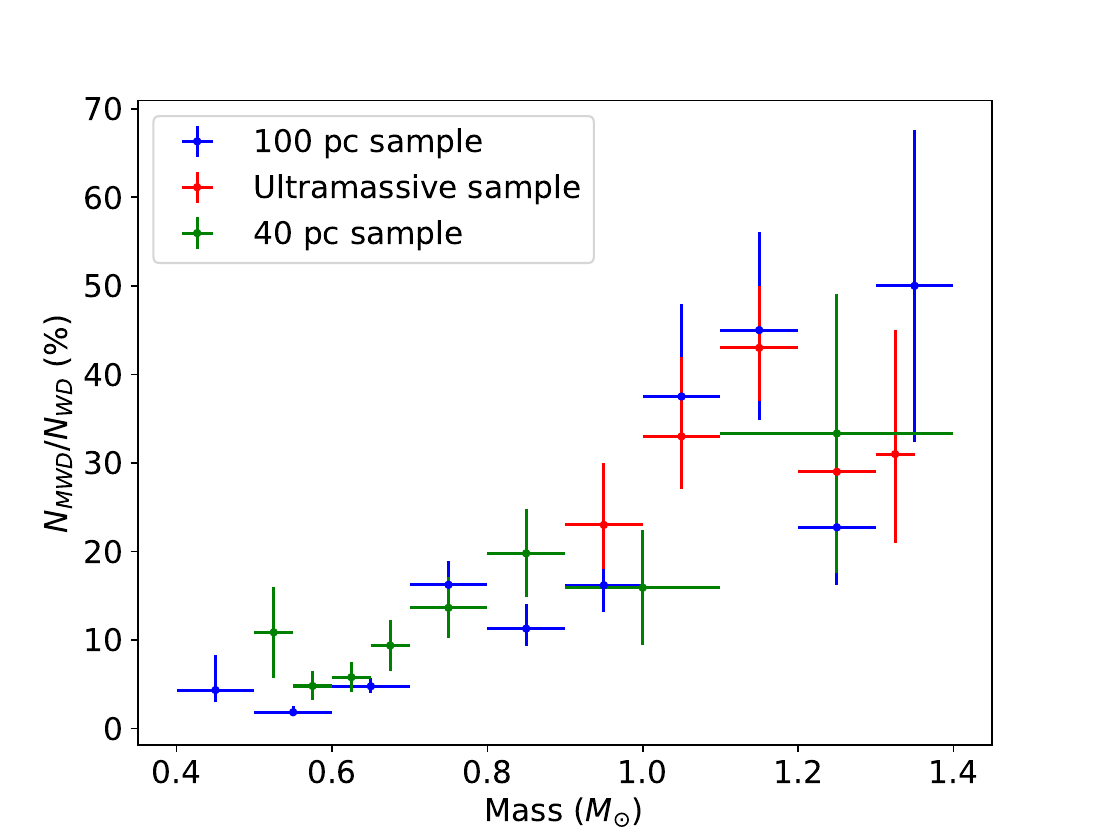}
    \includegraphics[width=3.75in, clip=true, trim=0.2in 0in 0in 0in]{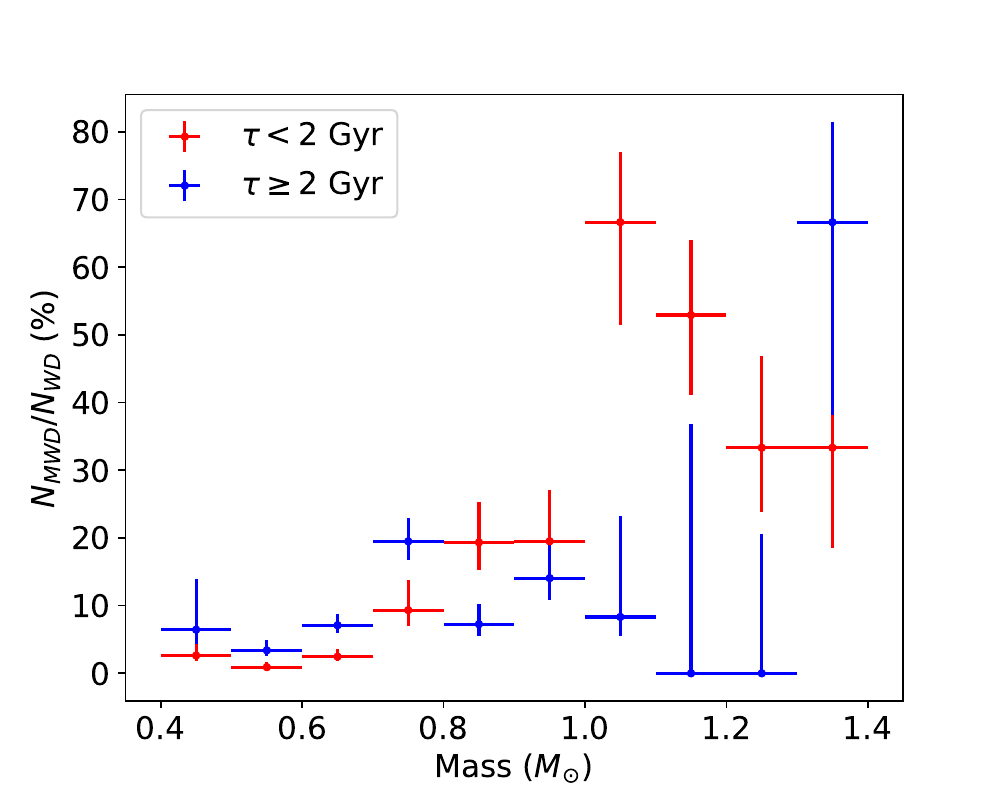}
    \caption{Top panel: the fraction of MWDs with H-dominated atmospheres in the SDSS 100 pc sample as a function of mass. Each mass bin is 0.1 $M_{\odot}$ wide. The trend in magnetic fraction is similar across each sample. Bottom panel: the fraction of MWDs with H-dominated atmospheres but separated by cooling age. Young objects are highly likely to be magnetic if they are massive, while older objects do not vary significantly in their magnetic fraction with mass.}
    \label{fig10} 
\end{figure}

In order to see how the fraction of MWDs changes with mass, we divide the sample into equal width bins and calculate the fraction of magnetic objects in each bin. We calculate the errors in the fraction using the binomial probability distribution discussed in \citet{Burgasser03}. The top panel of Figure \ref{fig10} shows the calculated magnetic fractions for H-dominated atmosphere WDs as a function of mass for our sample in addition to the ultramassive sample \citep{Jewett24} and the 40 pc sample \citep{Obrien24}. We see a clear increase in the magnetic fraction with mass. However, this initial picture is not entirely accurate as it mixes MWDs that originate from two distinct channels \citep{Bagnulo22}. It is perhaps more useful to analyze the magnetic fraction as a function of mass for different ages. In the bottom panel of Figure \ref{fig10} we calculate the magnetic fraction using the same mass bins but separating our sample into two age bins: objects that are either younger or older than 2 Gyr. We see that for the younger objects, the magnetic fraction quickly increases beyond 0.9 $M_{\odot}$. For the older objects, the magnetic fraction is relatively constant regardless of mass. In fact, the magnetic fraction below $\sim$0.9 $M_{\odot}$ is roughly the same regardless of age. Hence the increase in magnetic fraction with increasing mass as seen by \citet{Obrien24}, \citet{Jewett24}, and this work can be explained by the large fraction of MWDs among young ultramassive objects, which are likely products of binary evolution.

Given that we do not have spectropolarimetric observations for many of the cooler objects, our measurements present a firm lower limit. There can only be more MWDs if spectropolarimetric observations were available. While the 40 pc sample shows a similar decreasing trend in the magnetic fraction at lower mass, \citet{Obrien24} finds a higher fraction than our sample at these lower masses. This is likely due to their sample having a higher fraction of spectropolarimetry data, which become increasingly difficult to acquire with decreasing brightness (and thus distance). 

\subsection{Cooling Age}

\begin{figure}[!ht]
    \centering
    \includegraphics[width=3.5in, clip=true, trim=0in 0in 0in 0in]{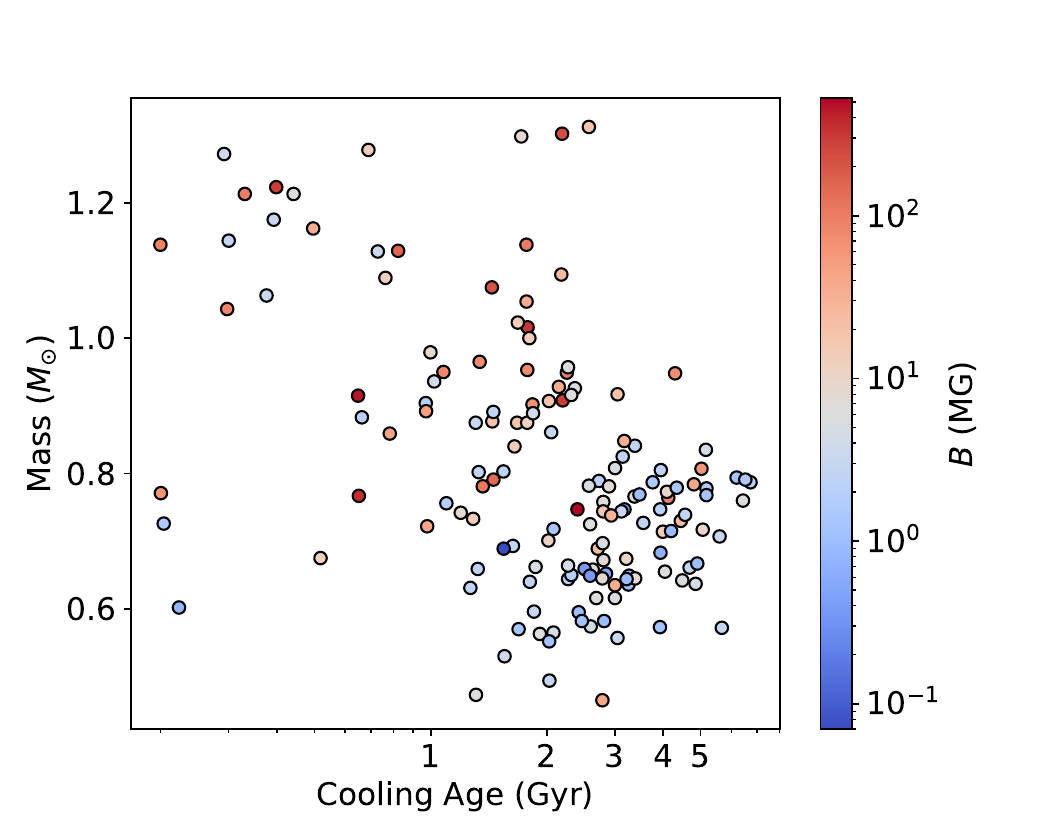}
    \caption{Mass as a function of cooling age colored by field strength. We see evidence of 2 distinct populations of MWDs, with lower mass objects possessing lower field strengths typically having longer cooling ages.} 
    \label{fig11}
\end{figure}

In Figure \ref{fig11} we plot mass and field strength as a function of cooling age to see if we detect a distinction in the sample based on cooling age. Interestingly, we do not detect any field strengths larger than 100 MG beyond a cooling age of $\sim$2.5 Gyr. The object with the strongest field, SDSSJ002128.60+150223.8, has a field strength of 527 MG \citep{Kulebi09} and a cooling age of 2.4 Gyr. In addition, we have 98 total objects with field strengths less than 10 MG, of which 54 (55\%) are older than 2.5 Gyr, and 67 (68\%) are older than 2 Gyr. We see that WDs older than $\sim$2 Gyr consistently have lower masses and weaker field strengths than those younger than 2 Gyr. While there are several objects at older ages with larger field strengths, these cases are the exception rather than the norm. 

\citet{Hardy23a} analyzed 651 objects labeled as magnetic or DAH in the Montreal White Dwarf Database, of which 140 targets could be well fitted with an offset dipole model. Similar to \citet{Amorim23}, there are significantly more MWDs that have not begun crystallization in their sample compared to ours (see their Figure 13). Again, this is expected as the \citet{Hardy23a} sample is also magnitude-limited such that the lower mass, hotter objects are more common.

\begin{figure}[!ht]
    \centering
    \includegraphics[width=3.75in, clip=true, trim=0in 0in 0in 0in]{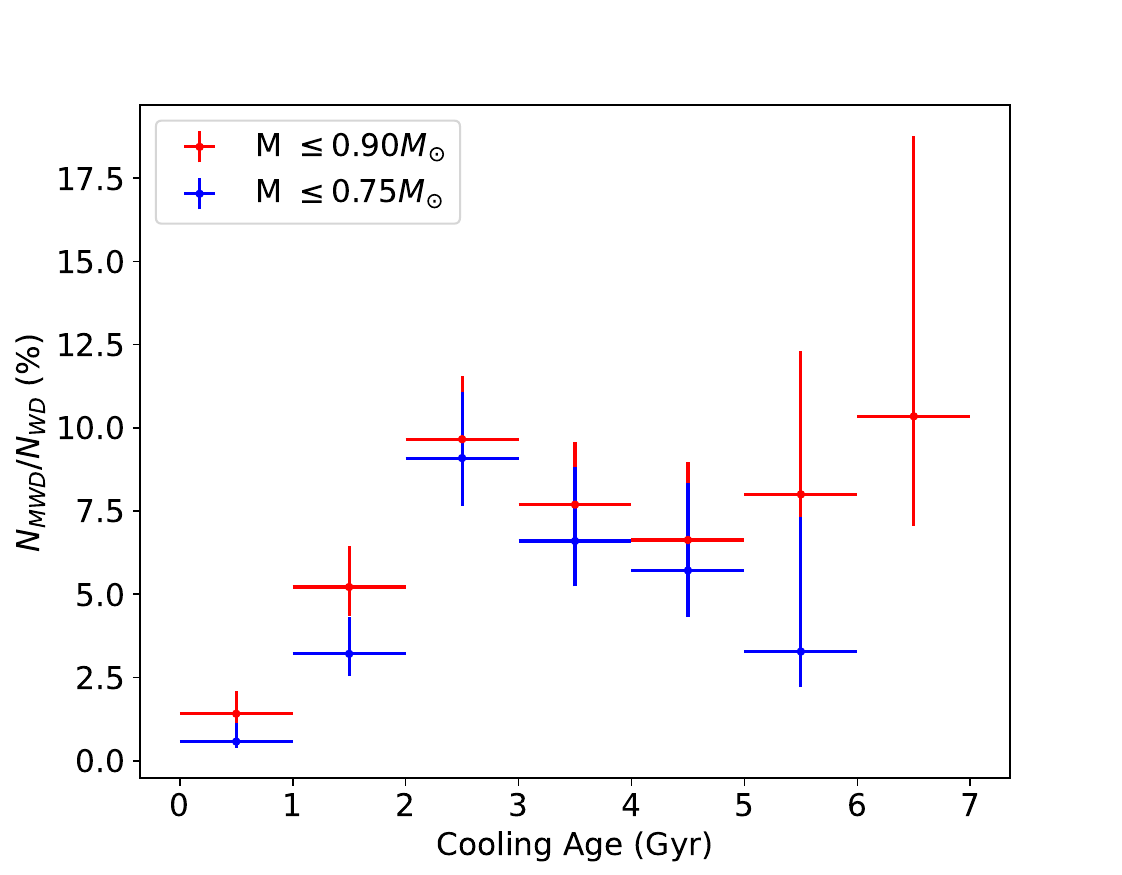}
    \caption{The fraction of MWDs as a function of cooling age. Each age bin is 1 Gyr wide. We see a clear increase in the magnetic fraction out to 2$-$3 Gyr, followed by a leveling off after 2$-$3 Gyr regardless of mass cutoff. Tighter constraints are needed for the oldest objects to verify the trend at older ages.}
    \label{fig12} 
\end{figure}

\citet{Bagnulo22} found an increase in the fraction of average-mass MWDs ($M <0.75 M_{\odot}$) with cooling age out to $\sim$4 Gyr, followed by a decline after $\sim$4$-$6 Gyr. These average-mass MWDs likely come from single-star evolution, and thus their magnetic fields likely come from crystallization or a previous stage of stellar evolution. Hence the observed variations in the magnetic fraction could indicate a slow relaxation of the field from the interior to the surface. We calculate the magnetic fraction as a function of cooling age shown in Figure \ref{fig12} using the same mass cutoff from \citep{Bagnulo22}. \citet{Jewett24} used 0.9 $M_{\odot}$ as the cutoff in their ultramassive sample since mergers contribute to a significant fraction of these WDs \citep{Temmink20}. Hence we also look at the trend in magnetic fraction as a function of cooling age for MWDs below 0.9 $M_{\odot}$. 

Similar to \citet{Bagnulo22}, we see an increase in the magnetic fraction with age. However, we find tighter error bars for the youngest objects given our larger sample size. The magnetic fraction in our sample peaks at a slightly younger age (2$-$3 Gyr), which could simply be due to our lack of spectropolarimetry and thus our inability to reliably detect fields in older objects. Beyond this peak, we see a leveling off in the magnetic fraction for objects below 0.75 $M_{\odot}$ and for objects less than 0.90 $M_{\odot}$. The decline in the fraction for older objects could be potentially due to ohmic decay, however it is more likely caused by the inherent difficulties of detecting fields in featureless WDs \citep{Bagnulo22}. The uncertainties for our 5$-$6 and 6$-$7 Gyr bins are also large given that we have only 6 and 3 MWDs in each bin respectively. Since magnetism occurs more frequently in higher mass WDs, the magnetic fraction for objects below 0.9 $M_{\odot}$ is larger at all ages than for those below 0.75 $M_{\odot}$, particularly at the oldest ages where we do not detect any MWDs below 0.75 $M_{\odot}$ and older than 6 Gyr.

While we do detect an increase in the incidence of magnetic fields up to 2$-$3 Gyr, we do not find a relation between the field strength and cooling age in the MWDs that have masses $< 0.75$ $M_{\odot}$. While lower-mass MWDs younger than 2 Gyr are relatively rare in our sample, the field strengths are generally between 1 and 10 MG. Beyond 2 Gyr, only one object has a field strength greater than 100 MG, and 11 are between 10 and 100 MG but at various cooling ages. \citet{Bagnulo22} found that both the frequency and strength of magnetic fields grow over time for these lower mass objects, with several objects possessing field strengths beyond 100 MG. It is again likely that the biases in our sample prevent us from obtaining a clearer picture of the evolution of these lower mass MWDs within our sample, as we are likely missing older objects that could possess notably stronger fields.

\subsection{Rotation Periods}

The fact that MWDs typically have shorter rotation periods than non-magnetic objects is well known \citep{Brinkworth13,Kawaler15,Hermes17,Kawka20,Hernandez24,Oliveira24}. A merger remnant from two CO WDs is predicted to rotate on the order of minutes \citep{Schwab21}, while crystallization could induce a dynamo independent of the WD's rotation rate during efficient convective mixing \citep{Fuentes24}. Hence any known periods in our sample could lend insight into the mechanism that generates the magnetic fields. 

We search for photometric variability in our sample using the Transiting Exoplanet Survey Satellite (TESS) and Zwicky Transient Facility (ZTF). For the TESS data, we check the 20 sec and 2 min cadence light curves available on the Mikulski Archive for Space Telescopes (MAST), as well as the Full Frame Images (FFI) for targets that do not have short cadence light curves available. We use TESSExtractor \citep{Brasseur19,Serna21} to extract the light curves from the FFIs. We generate Lomb-Scargle periodograms to determine if there is a strong signal in each light curve, and we report the photometric rotation periods only if our results satisfy multiple criteria: the signal must be above the 1\% False Alarm Probability; the signal occurs in multiple TESS sectors or in ZTF; there is little aliasing or any potential aliasing can be explained as harmonics of the true signal; and the period for a given object must closely match values reported in the literature. We also report the rotation period if it has been previously obtained through a separate method such as spectropolarimetry or time-series spectroscopy. 

\begin{figure}[!ht]
    \centering
    \includegraphics[width=3.5in, clip=true, trim=1.3in 2.75in 0.5in 3in]{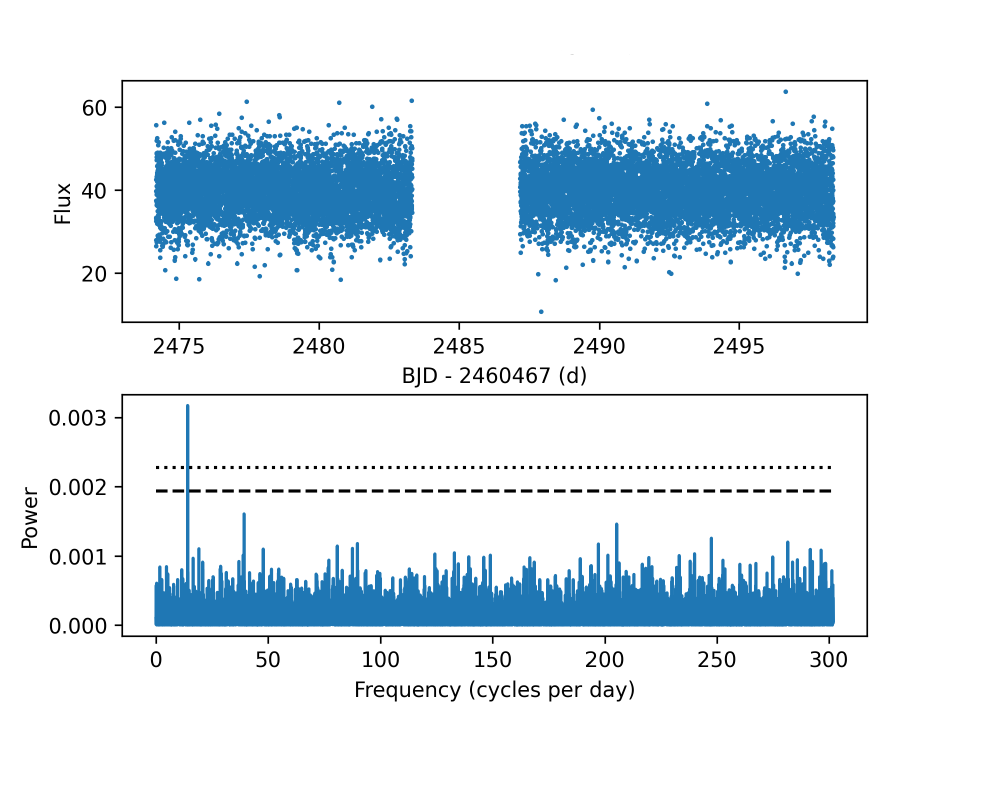}
    \caption{TESS light curve and Lomb-Scargle for J0153+1808. The dashed and dotted lines represent the 10\% and 1\% False-Alarm Probabilities. We detect a significant signal in two sectors at 1.69 hours, which matches the result from \citet{Oliveira24}.} 
    \label{fig13}
\end{figure}

\begin{figure}[!ht]
    \centering
    \includegraphics[width=3.5in, clip=true, trim=1.3in 2.2in 0.5in 2.5in]{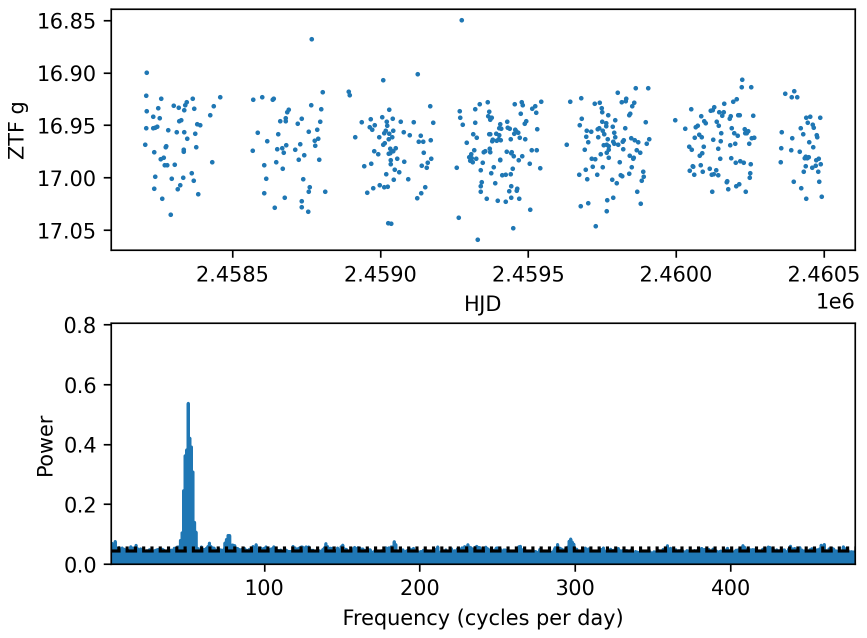}
    \caption{ZTF light curve and Lomb-Scargle for J1852+1833. There  is a significant signal at 0.48 hours.
    %Bottom panel: trailed spectra of J1852+1833 taken at APO with 10 minute exposures. Here we do not see clear line position variations, but instead the target is significantly fainter for the first several exposures, leading to noisier spectra.
    } 
    \label{fig14}
\end{figure}

Figure \ref{fig13} shows an example TESS light curve and periodogram for a variable DAH, SDSSJ015342.00+180857.9 (J0153+1808). There is a significant signal at 1.69 hours in 2 TESS sectors, and \citep{Oliveira24} obtains the same period. While we do not detect a signal for this object in ZTF, there is little data available so this is not surprising. Hence we adopt the 1.69 hour value. Figure \ref{fig14} shows the ZTF light curve and periodogram for SDSSJ185218.63+183355.4 (J1852+1833). This object shows significant variability both in ZTF and TESS but at different periods. The ZTF light curve favors a very short rotation period of 0.48 hours, while TESS shows convincing evidence of a 6.7 hour period. This target lies in a particularly dense field with many nearby sources. Given the large pixel sizes in TESS that leads to source blending, we adopt the ZTF signal.

Of our 163 MWDs, we have confident rotation period measurements for 26 of them (16\%). We report these values in Table \ref{tab2}. We do not detect significant variability in the objects with only TESS FFIs available. Many of our objects have been previously identified by \citet{Oliveira24} who conducted a wide TESS search of variable MWDs. Our fraction of photometrically variable targets is slightly lower than the 20\% obtained for the 20 pc volume by \citet{Farihi24}. One object from the literature we exclude is SDSSJ125715.54+341439.4, which does not show significant variability in our TESS light curves, while \citet{Oliveira24} found a period of 0.092 hours. They only detected variability in one sector for this target, so we do not include it here. We also exclude SDSSJ133250.73+011706.3, which has confirmed spectroscopic variations \citep{Moss23} but an unconstrained period. We do not detect photometric variability in this target, and there is no reported value in the literature. 

\begin{deluxetable}{ccc}
    \tabletypesize{\scriptsize}
    \tablecolumns{3} \tablewidth{0pt}
    \centering
    \caption{\label{tab2} Known rotation periods in our sample.}
    \tablehead{\colhead{Object} & \colhead{Period (hrs)} & \colhead{Source}\\}
    \startdata
        J0006+3104 & 0.386 & TESS, ZTF, \citet{Oliveira24}\\
        J0050-0326 & 0.672 & \citet{Jewett24,Oliveira24}\\
        J0150+2835 & 0.288 & ZTF\\
        J0153+1808 & 1.69 & TESS, \citet{Oliveira24}\\
        J0718+3731 & 0.188 & TESS, ZTF, \citet{Oliveira24}\\
        J0748+1125 & 0.192 & ZTF\\
        J0835+5332 & 2.16 & \citet{Steen24}\\
        J0856+1611 & 5.70 & TESS, ZTF, \citet{Hermes17,Oliveira24}\\
        J1018+0111 & 1.65 & TESS, ZTF, \citet{Hernandez24}\\
        J1029+1127 & 0.681 & TESS, ZTF, \citet{Moss23,Oliveira24}\\
        J1039-0325 & 0.501 & TESS, \citet{Oliveira24}\\
        J1105+5225 & 1.92 & ZTF\\
        J1516+2803 & 0.576 & \citet{Moss23}\\
        J1543+3021 & 1.31 & TESS, ZTF, \citet{Oliveira24}\\
        J1621+0432 & 0.672 & \citet{Bagnulo24}\\
        J1628+2332 & 5.3 & TESS\\
        J1630+2724 & 14.86 & ZTF\\
        J1659+4401 & 0.704 & \citet{Jewett24,Oliveira24}\\
        J1707+3532 & 0.576 & ZTF, \citet{Jewett24}\\
        J1852+1833 & 0.480 & ZTF\\
        J1956-0102 & 34.78 & TESS, \citet{Valyavin11}\\
        J2211+1136 & 0.0192 & \citet{Kilic21}\\
        J2223+2319 & 1.42 & ZTF\\
        J2257+0755 & 0.37 & ZTF, \citet{Williams22}\\
        J2318+1236 & 428.54 & \citet{Schmidt91}\\
        J2332+2658 & 2.77 & \citet{Brinkworth13}\\   
    \enddata    
\end{deluxetable}

\begin{figure}[!ht]
    \centering
    \includegraphics[width=3.5in, clip=true, trim=0in 0in 0in 0in]{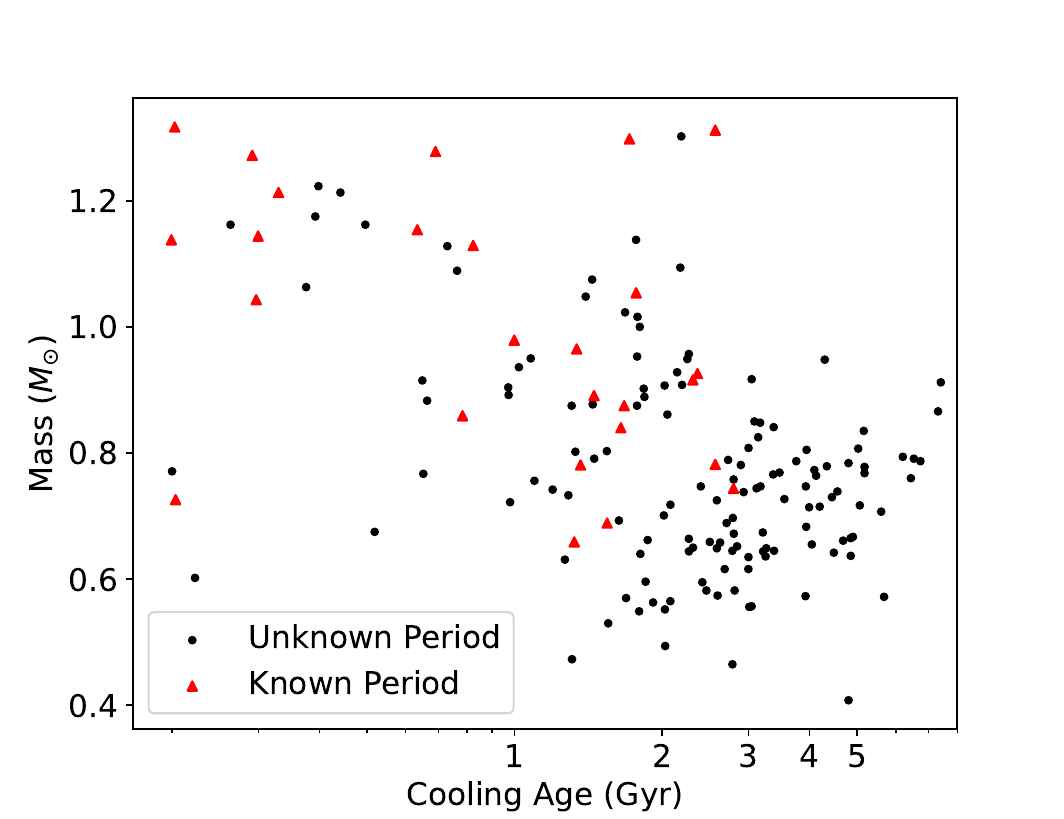}
    \caption{Mass-age distribution of our sample with the objects that have known rotation periods marked with red triangles. The majority of our objects with known periods are young and high mass.} 
    \label{fig15}
\end{figure}

The median period for our 26 objects is 0.69 hours. This result is notably shorter than the results found by \citet{Hernandez24} and \citet{Oliveira24}, who found median values of 6.2 and 3.5 hours respectively for their TESS samples. We do not detect strong trends in the period with either mass or field strength, similar to previous work \citep{Brinkworth13,Jewett24,Hernandez24,Oliveira24}. However, we do see that the objects with known periods tend to have higher masses and lower cooling ages as shown in Figure \ref{fig15}. Of the 26 objects with known periods, only five are older than 2 Gyr. The median mass and field strength for our variable objects is 0.97$M_{\odot}$ and 15 MG respectively. Since these variable objects tend to have higher mass and field strengths than the entire sample, in addition to having short rotation periods, photometric variability tends to be more common among likely merger remnants.

\citet{Bagnulo24} found that 16 of 17 normal-mass MWDs older than 2 Gyrs and possess strong fields ($B \geq 10$ MG) are non-variable. However, they found that 4 of 8 MWDs in the same mass and age range but with intermediate field strengths (1 $< B <$ 10 MG) are non-variable. Of our 89 targets that are older than 2 Gyr and have field measurements, only 23 have fields stronger than 10 MG, and 57 of our old targets have intermediate field strengths, of which a sizable fraction should have variable fields. Since we do not conduct spectropolarimetry, we are unable to determine rotation periods for many of these older objects.

\subsection{Tangential Velocities}

High tangential velocities ($v_{\rm tan} > 50$ km/s, \citealt{Wegg12}) can be an indicator of a merger event if the WD's cooling age is fairly young. Velocity dispersion increases with age \citep{Nordstrom04}, but a merger event resets the cooling age, making the product appear younger than its actual age. However, using just the velocity to identify mergers is difficult given that not all likely merger products have high velocities. \citet{Wegg12} found that the velocity dispersion of WDs decreases with mass, and concluded that their high-mass ($M> 0.95 M_{\odot}$) targets came from single-star evolution given that none of these targets had high tangential velocities. But \citet{Cheng20} found that $\sim$30\% of their targets with masses $> 1 M_{\odot}$ come from mergers. Among the ultramassive WDs in the 100 pc sample, \citet{Jewett24} found that only five out of 66 MWDs had high velocities, despite classifying the remaining 61 as likely merger remnants.

\begin{figure}[!ht]
    \centering
    \includegraphics[width=3.5in, clip=true, trim=0in 0in 0in 0in]{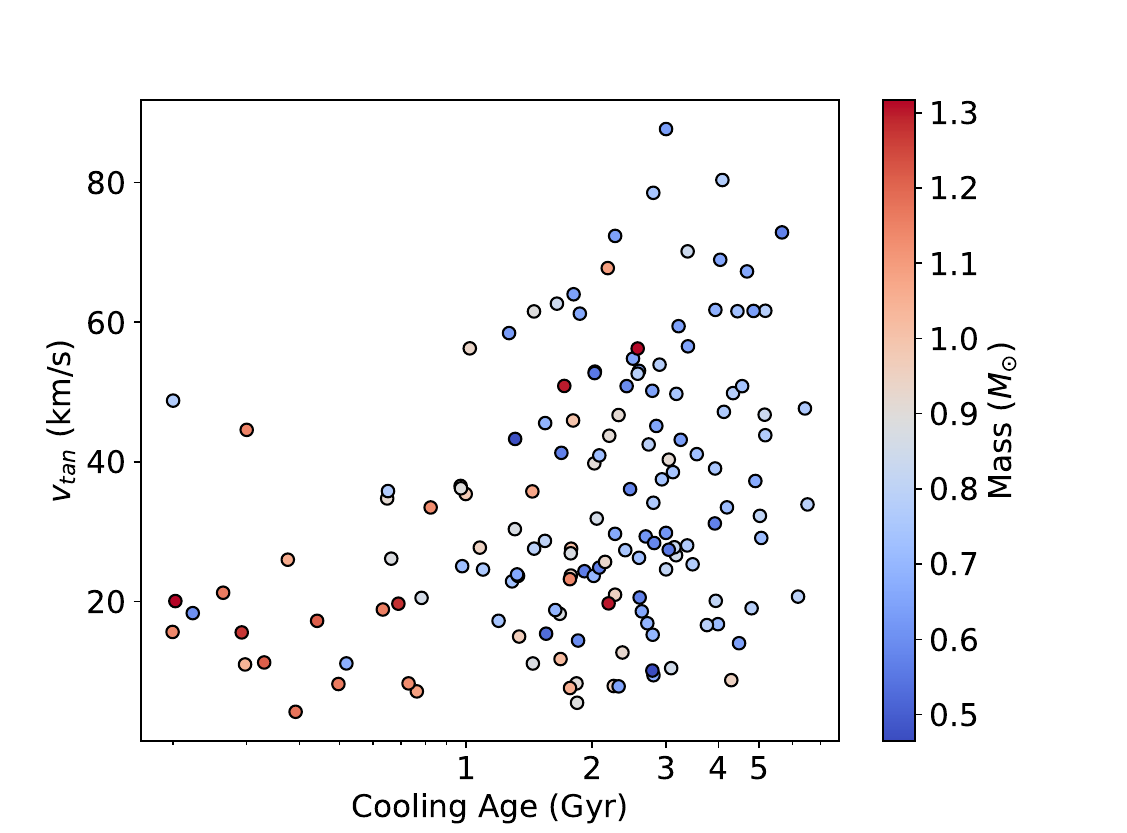}
    \caption{Tangential velocity distribution for the DAHs in our sample. The high-mass MWDs in our sample have velocities almost exclusively below 50 km/s, while more average mass and older MWDs show a large dispersion in their velocities.} 
    \label{fig16}
\end{figure}

To determine if our magnetic objects are kinematic outliers, we calculate the tangential velocities of our sample using the Gaia DR3 proper motions and parallaxes. Figure \ref{fig16} shows the tangential velocities for the DAHs in our sample. The velocity dispersion increases with age as expected, with few targets possessing high velocities before 2 Gyr. Interestingly, our high mass targets have some of the lowest velocities in the sample despite being likely merger remnants based on their masses and field strengths. Of the 26 objects with known rotation periods, 23 have periods shorter than 6 hrs. Of these 23 fast-spinning objects, 15 have masses $> 0.9 M_{\odot}$. These 15 objects are thus likely merger remnants. For these 15 objects, only four have tangential velocities $> 50$ km/s. Hence not all merger remnants can be identified based on their kinematics alone.

\section{Evidence of Two Populations} \label{stats}
\subsection{Kolmogorov-Smirnov Tests}

The primary goal of this study is to test the two population hypothesis initially proposed by \citet{Valyavin99} and confirmed by \citet{Bagnulo22} with a large, volume-limited survey. In Figure \ref{fig11}, we can already see evidence of these two populations: young objects tend to be more massive and have stronger fields, while older objects are less massive and have weaker fields. These populations have distinct masses and field strengths, which indicate two separate formation channels that lead to different MWDs. To investigate these channels, we split the magnetic objects into two groups based on cooling age: those that are older and younger than 2 Gyr. 

While the exact age cutoff we choose is somewhat flexible as we will show, the important distinction is that one group is comprised of relatively young MWDs, while the other group contains older objects. \citet{Bagnulo22} found that MWDs older than 2 Gyr have similar masses to those of non-MWDs, and that the lower mass WDs do not show evidence of magnetism until roughly 2 Gyr in cooling age. In addition, \citet{Bagnulo24} found that MWDs across all masses and older than 2 Gyr with high field strengths ($B \geq 10$ MG) show little polarimetric variability, which could be an indicator of their field origin. Hence we conclude that a 2 Gyr age cutoff is sufficient to study the possible separate formation channels in these MWDs.

\begin{figure}[!ht]
    \centering
    \includegraphics[width=3.5in, clip=true, trim=0in 0in 0in 0in]{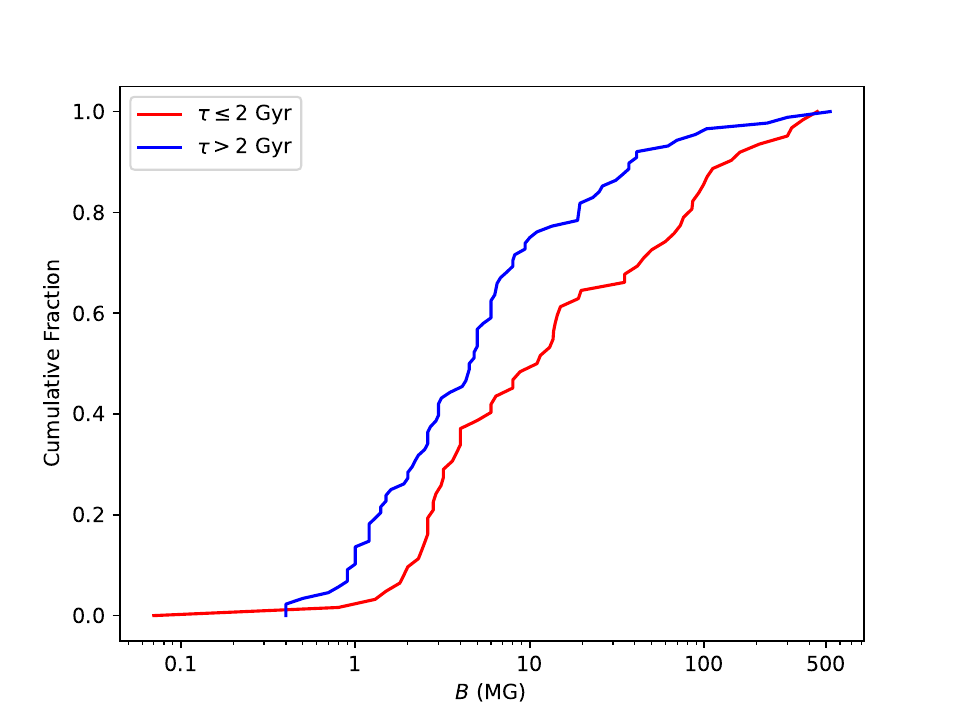}
    \caption{Cumulative distribution of magnetic field strengths for the old ($\tau > 2$ Gyr, blue line) population and young ($\tau \leq 2$ Gyr, red line) population. There is a clear separation between the two distributions, with younger objects possessing stronger fields than older objects.} 
    \label{fig17}
\end{figure}

We first perform nonparametric two-sample Kolmogorov-Smirnov (K-S) tests \citep{Kolm33,Smirnov39} on the field strengths and masses of these two age groups. These tests are utilized to determine if two samples, in our case objects younger and older than 2 Gyr, belong to the same underlying distribution. The statistical significance we provide here is in the form of a two-tailed p-value, and we compare it to a significance level of $\alpha = 0.05$ to show evidence against our null hypothesis, which is that any MWDs younger or older than 2 Gyr belong to the same population. Figure \ref{fig17} shows the cumulative distribution of the field strengths for each age group. We see a clear difference in the field strengths based on cooling age, with younger objects clearly possessing stronger fields than older objects. Performing a two-sample K-S test on these separate field strength groups yields a p-value of 0.01. Hence we reject the null hypothesis that the field strengths in our two samples originate from the same parent distribution. 

The discrepancy in field strength between the two age groups could partly be explained by the biases in our sample. Average mass MWDs are expected to have weak fields at young ages which then grow over time \citep{Bagnulo22}. In general, we are not sensitive to magnetic fields below $\sim$1 MG, so many of the hot, average-mass WDs we classify as non-magnetic could simply have weak fields. Additionally, as these WDs cool off sufficiently, their spectral lines vanish so we would be unable to detect the potential strong fields at old ages. Hence the $\tau \leq 2$ Gyr distribution could be missing weak fields, while the $\tau > 2$ Gyr distribution could be missing strong fields.

However, the actual contributions from these sources is likely to be insignificant for two reasons. The first is that even when using detection methods that can probe fields at kG strength, the young and average-mass MWDs are quite rare. \citet{Bagnulo22} found that only 2.5\% or less of WDs with mass $< 0.75$ $M_{\odot}$ and $\tau < 0.6$ Gyr are magnetic. Even out to 2 Gyr in the 20 pc volume, there are are only 5 more MWDs in this mass range compared to $\sim$20 non-magnetic WDs. Hence it is highly unlikely that the young MWD group has significant contribution from weak fields even if we cannot detect them in this sample. The second reason is that even though our low-resolution spectroscopy is not sensitive to fields in the coolest WDs where H lines disappear, 
 we can still reliably detect fairly strong fields down to $\sim$6000 K (see below).

\begin{figure}[!ht]
    \centering
    \includegraphics[width=3.5in, clip=true, trim=0in 0in 0in 0in]{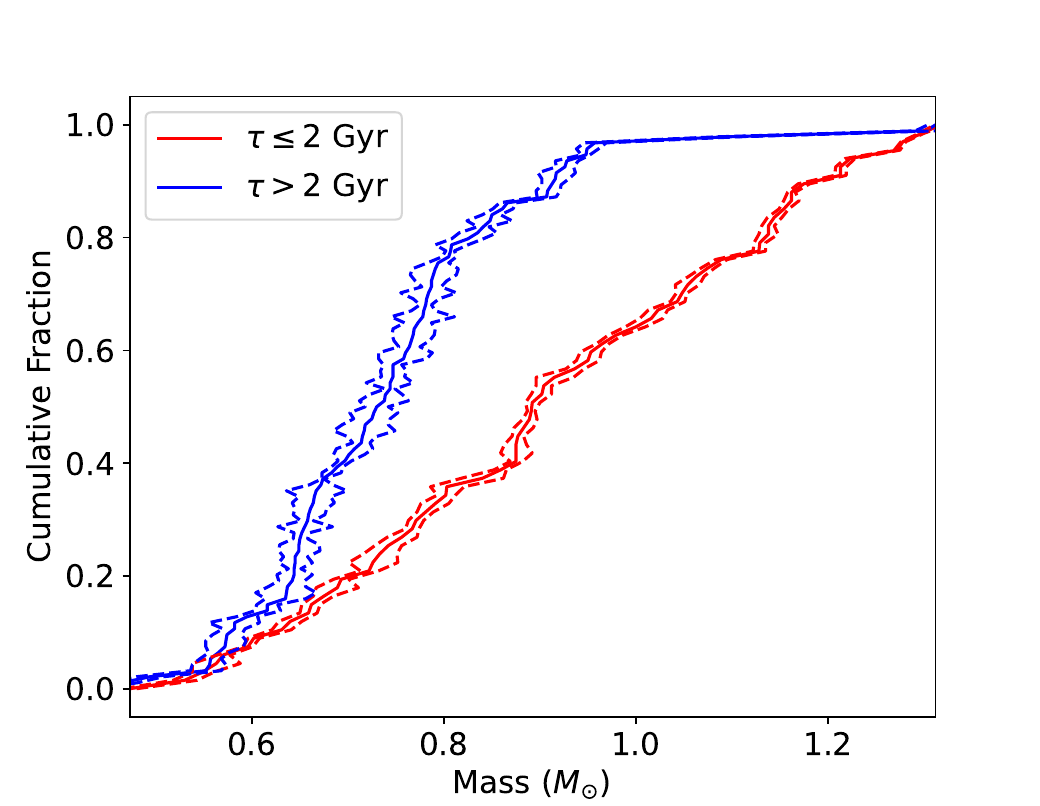}
    \includegraphics[width=3.5in, clip=true, trim=0in 0in 0in 0in]{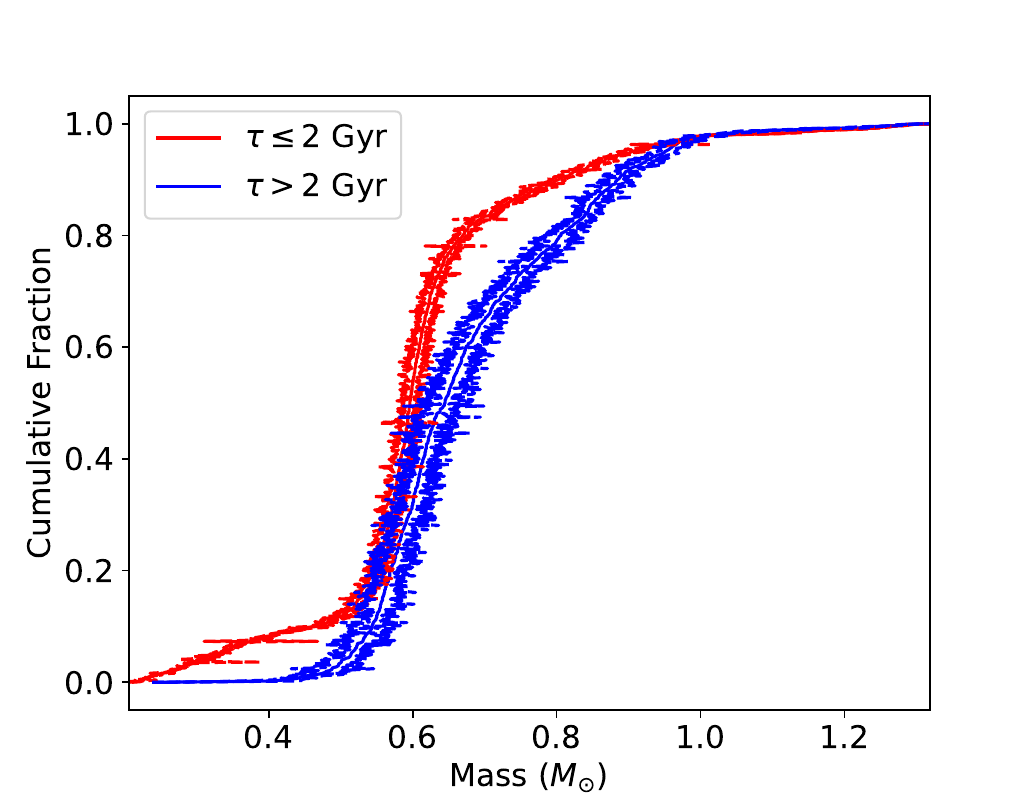}
    \caption{Cumulative distribution of stellar masses separated by cooling age for the MWDs (top panel) and the non-magnetic DA WDs (bottom panel). The dashed lines show the upper and lower mass limits from our fits. Errors of less than 5\% are extremely common. Older MWDs tend to have lower masses than younger MWDs, while the non-magnetic DAs show the opposite trend.}
    \label{fig18}
\end{figure}

Mergers are expected to produce some of the strongest fields among MWDs, and these fields should instantly appear at the surface \citep{Garcia12}. Naturally, a group of WDs that originate from mergers should also be more massive than those that come from single-star evolution \citep{Temmink20}. Figure \ref{fig18} shows the cumulative distributions for the masses using our 2 Gyr age cutoff for both the magnetic and non-magnetic objects in the 100 pc sample. The young magnetic targets are notably more massive than the older targets, while the distributions of the non-magnetic targets seem to show the opposite. Performing the K-S tests for the masses of the magnetic and non-magnetic objects yields p-values of $2.3\times10^{-8}$ and $7.7\times10^{-18}$ respectively, so we again reject the null hypothesis. For both field strength and mass separated by younger and older MWDs, the two-sample K-S test has provided evidence with 95\% certainty that these two groups are observationally distinct groups of MWDs.

While we choose a 2 Gyr age cutoff in these tests, we can achieve similar statistical significance for a small range of cooling ages. We perform multiple K-S tests with age cutoffs every 0.5 Gyr ranging from 0.5 to 4.0 Gyr. While the difference in mass among the MWDs is statistically significant regardless of what age we choose, we only obtain p-values less than 0.05 between 1.0 and 3.5 Gyr for the field strengths, with our lowest p-value occurring at 2.5 Gyr. 

\subsection{A Gaussian Mixture Model Approach}
To further confirm the existence of two distinct populations in our sample, we use a Gaussian mixture model (GMM) with 2 components to group objects together based on similar parameters \citep{Day69,Mclachlan00}. A GMM is an algorithm that calculates the parameters to a set of multivariate Gaussian distributions, such that the likelihood of data being described by those distributions is maximized. In this case, this data is the set of magnetic field strengths, cooling ages, and masses observed, and we want to show that this data can be described by two Gaussian distributions in a 3-dimensional space. In this way, we are looking to find evidence of these two populations being more distinct in higher-dimensional space. We make this calculation using \texttt{scikit-learn}'s \texttt{GaussianMixture} model \citep{scikit}, which provides the optimal values of means and covariance matrices for the two Gaussian distributions that describe our groups. This method is advantageous for our sample as there is clear overlap between the two visually proposed groups, which necessitates the use of a clustering method based on probability as opposed to direct assignment (e.g. $k$-means clustering). With GMMs, a probability of each object belonging to either group can be determined, and therefore group assignment can be decided based on the group with a higher probability.

\begin{figure}[!ht]
    \centering
    \includegraphics[width=3.5in, clip=true, trim=0in 0in 0in 0in]{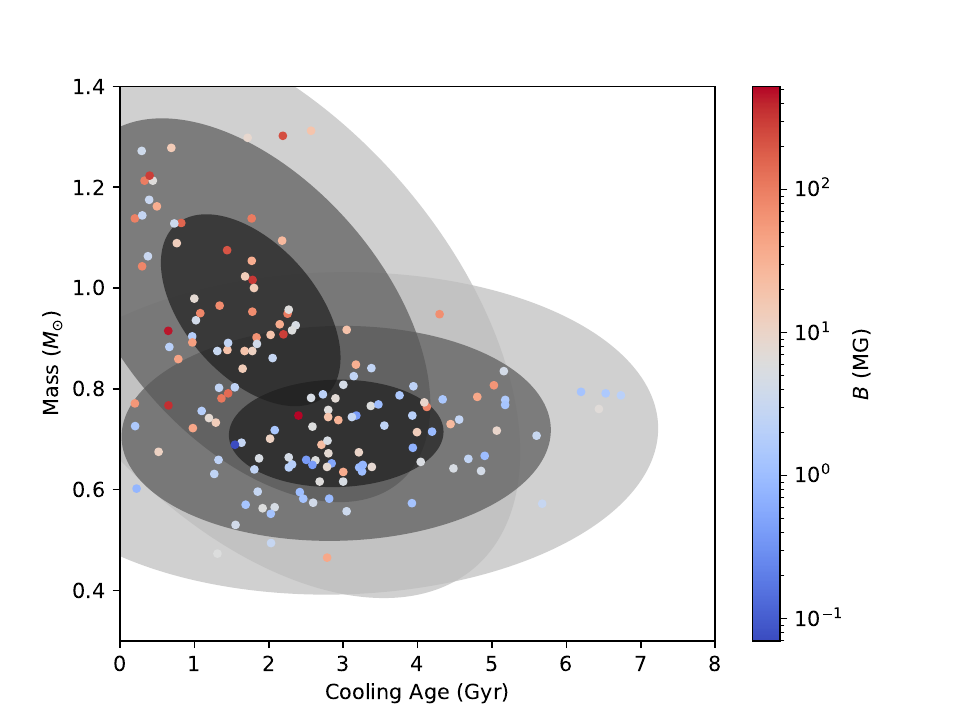}
    \caption{Similar to Figure \ref{fig11} but including the clusters from our GMM analysis. The shaded regions represent the 1, 2, and 3$\sigma$ intervals from dark to light gray respectively.}
    \label{fig19}
\end{figure}

Figure \ref{fig19} shows the calculated clusters and a visual representation of 1-, 2-, and 3$\sigma$ intervals projected onto the axes of the figure given by the covariance matrix of the GMM \citep[see][for more details]{Burrow20} The algorithm detects two groups at low and high mass (and correspondingly, old/young age) with little overlap at the 1$\sigma$ level. The center of each cluster occurs at 2.9 Gyr, 0.71 $M_{\odot}$, 3.7 MG and 1.8 Gyr, 0.96 $M_{\odot}$, 84 MG respectively. This result combined with the K-S tests provide clear and strong evidence that there exists two formation channels that lead to magnetic dynamos at different evolution stages and field strengths based on our sample. It is important to note that these clusters represent only the MWDs in our 100 pc sample, and not necessarily the true distribution of MWDs. The GMM cannot account for the young, average mass, weak field objects nor the old, average mass, strong field objects that we are likely missing. 

\begin{figure}[!ht]
    \centering
    \includegraphics[width=3.5in, clip=true, trim=0in 0in 0in 0in]{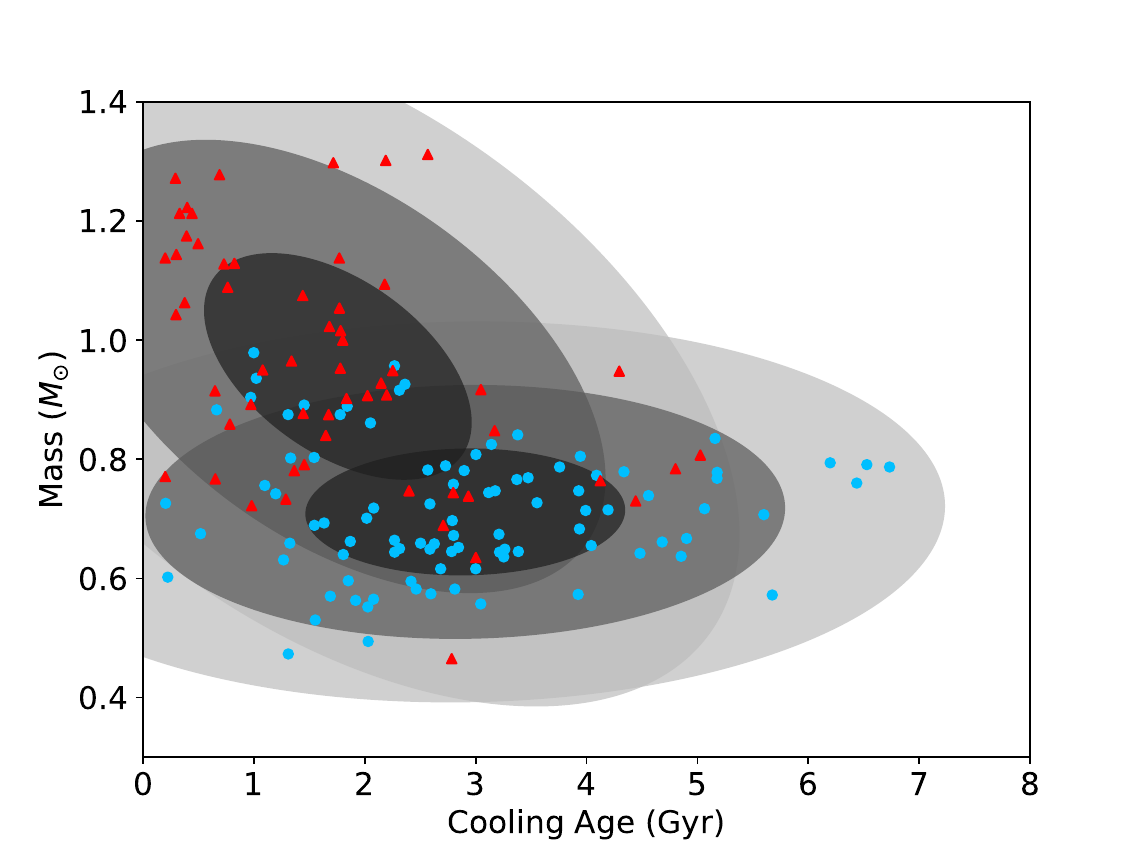}
    \caption{Mass as a function of cooling age with each point marked by which cluster the GMM method assigns them to. Red points are assigned to the young, high mass, strong field cluster while blue points are assigned to the old, low mass, weak field cluster.}
    \label{fig20}
\end{figure}

Figure \ref{fig20} shows the same mass-cooling age plot but with the points marked by which cluster the GMM assigns each point to. For example, there are 3 points at $\sim$0.8 $M_{\odot}$ and $\sim$5 Gyr, but they have field strengths larger than 10 MG, so the algorithm classifies them as belonging to the young population. These ``outlier'' targets could realistically belong to either group, so confirming the formation channel in these is naturally difficult. However, looking at the sample holistically, the measurements suggest that two formation channels lead to two distinct groups of MWDs. 

\section{Discussion}

We provide further evidence of the existence of two distinct MWD populations likely created by two distinct formation mechanisms. These populations were initially proposed by \citet{Valyavin99} based on $T_{\rm eff}$ and mass estimates for 49 and 35 MWDs, respectively. They found that the incidence of magnetism is approximately seven times higher in WDs with mass $> 1.1$ $M_{\odot}$ than in WDs below this mass. Additionally, 70\% of these ultramassive WDs have field strengths $> 100$ MG, while none of the MWDs below this mass possess such high field strengths. Hence \citet{Valyavin99} predicted that these ultramassive highly magnetic WDs come from binary evolution.

\citet{Bagnulo22} confirmed these populations based on the analysis of all WDs within the 20 pc volume (see \citealt{Bagnulo21}) and WDs younger than 0.6 Gyr within 40 pc. The authors made several key findings, including a clear increase over time in the magnetic fraction of WDs with mass $< 0.75$ $M_{\odot}$. Additionally, among the WDs younger than 0.6 Gyr, $\sim$85\% with masses $> 1.1$ $M_{\odot}$ are magnetic compared to $\lesssim2.5$\% for masses $< 0.75$ $M_{\odot}$. Their mass distribution also provides important insights: the average mass of their young MWDs is 0.98 $M_{\odot}$ compared to 0.65 $M_{\odot}$ for objects older than 2 Gyr. The high-mass MWDs have high field strengths on the order of MG while comparable fields do not appear in the lower mass MWDs until after 2 Gyr of cooling. We now discuss these two populations with respect to our 100 pc sample.

\subsection{Biases in our Sample}

The low-resolution spectra from SDSS leads to difficulties in detecting magnetic objects with field strengths weaker than $\sim$1 MG \citep{Kepler13}, as well as in older WDs that lose their spectral features as they reach $T_{\rm eff}\lesssim$11,000 K for He-dominated atmospheres and $\lesssim$5,000 K for H-dominated atmospheres. We can analyze these biases and their effects on our sample with regards to the two magnetic populations put forth by \citet{Valyavin99} and \citet{Bagnulo22}. The young, massive, strong MWDs should not be significantly affected by these biases. Even though strong field strengths can wipe out absorption features ($B \gtrsim 100$ MG), these objects are almost exclusively found at hotter temperatures, such that we can infer the presence of a strong magnetic field if no absorption lines are detected. Polarimetric measurements would certainly aid in further constraining the field strength, but otherwise are not needed to detect a field for these targets.

The group of average mass, older, weakly magnetic objects is a different story. \citet{Bagnulo22} found that there is an evolutionary trend in the average mass MWD population. At young ages, magnetism is rare and the fields that are present are fairly weak. For WDs younger than 0.6 Gyr, less massive than 0.75 $M_{\odot}$ and within 40 pc, $\lesssim2.5$\% of them are magnetic. In addition, the few MWDs within this age and mass range have field strengths less than 1 MG. Hence while these young, average mass MWDs are quite rare, we would likely not detect them given our bias against such weak fields. At older ages, both the frequency of MWDs and strength of their fields increase. The first average mass MWD with a MG field does not appear until approximately 2 Gyr, but 11 of 64 MWDs older than 3 Gyr and $M \leq 0.75$ $M_{\odot}$ have field strengths comparable to the young, high mass population. While we do detect MWDs beyond 3 Gyr, it certainly does become more challenging. Thus it is almost guaranteed that we are underestimating the number of average mass MWDs both at very young ages where the fields are too weak to detect and at older ages where spectral lines vanish.

\begin{figure*}
    \centering
    \includegraphics[width=0.33\textwidth,clip=true,trim=0.5in 1in 0.5in 0in]{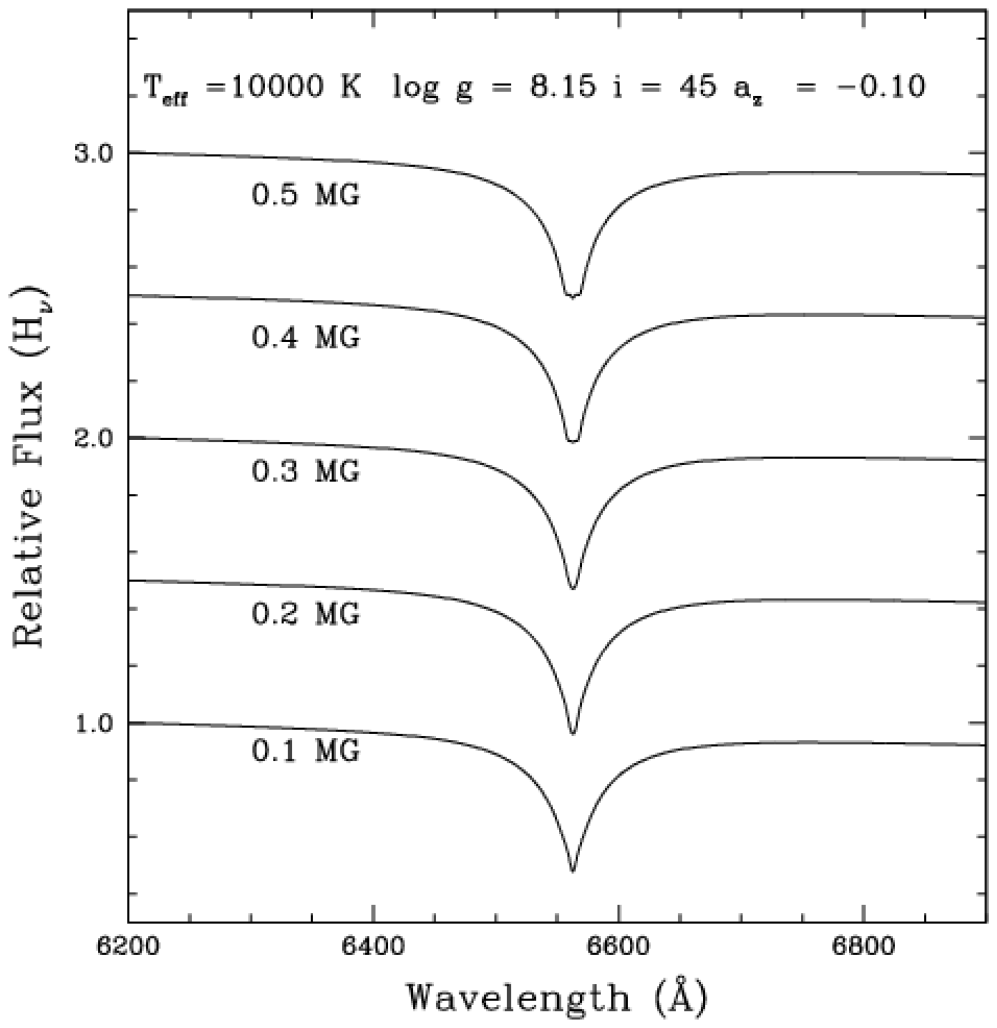}
    \includegraphics[width=0.33\textwidth,clip=true,trim=0.5in 1in 0.5in 0in]{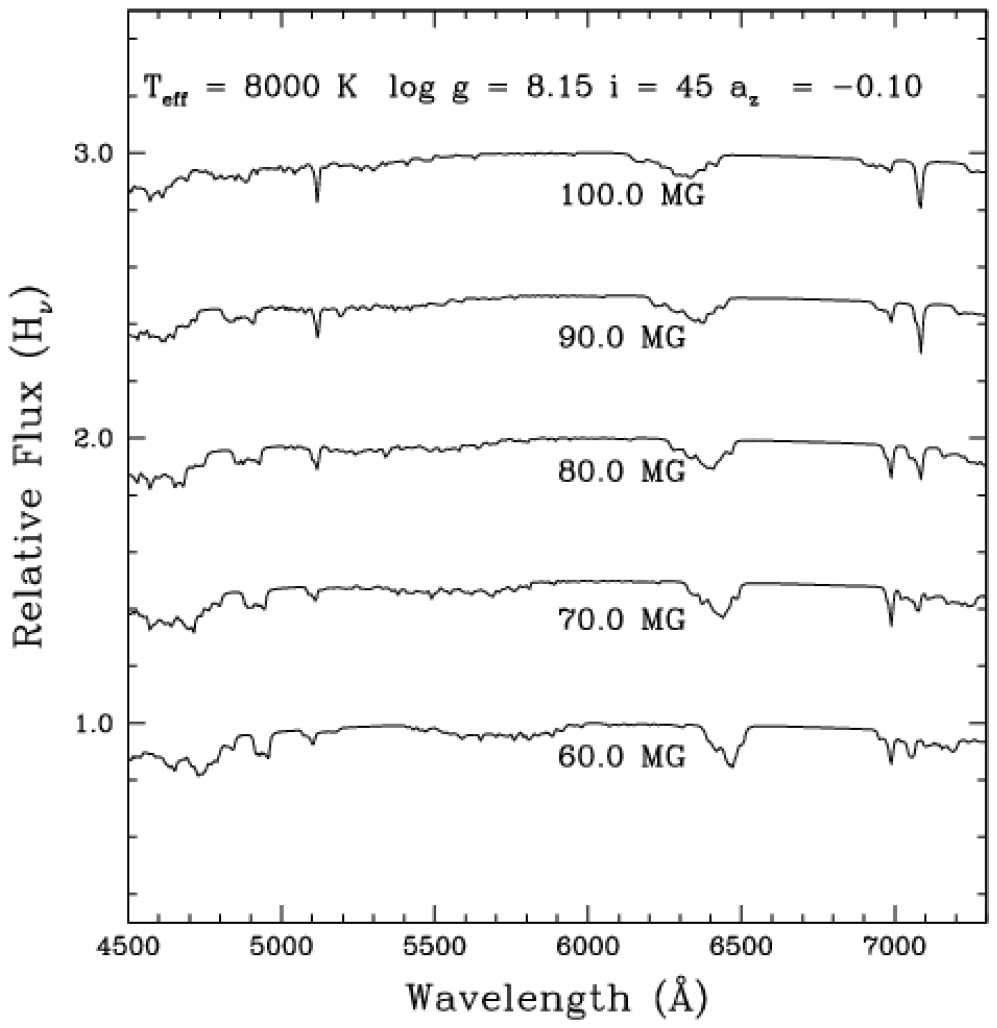}
    \includegraphics[width=0.33\textwidth,clip=true,trim=0.5in 1in 0.5in 0in]{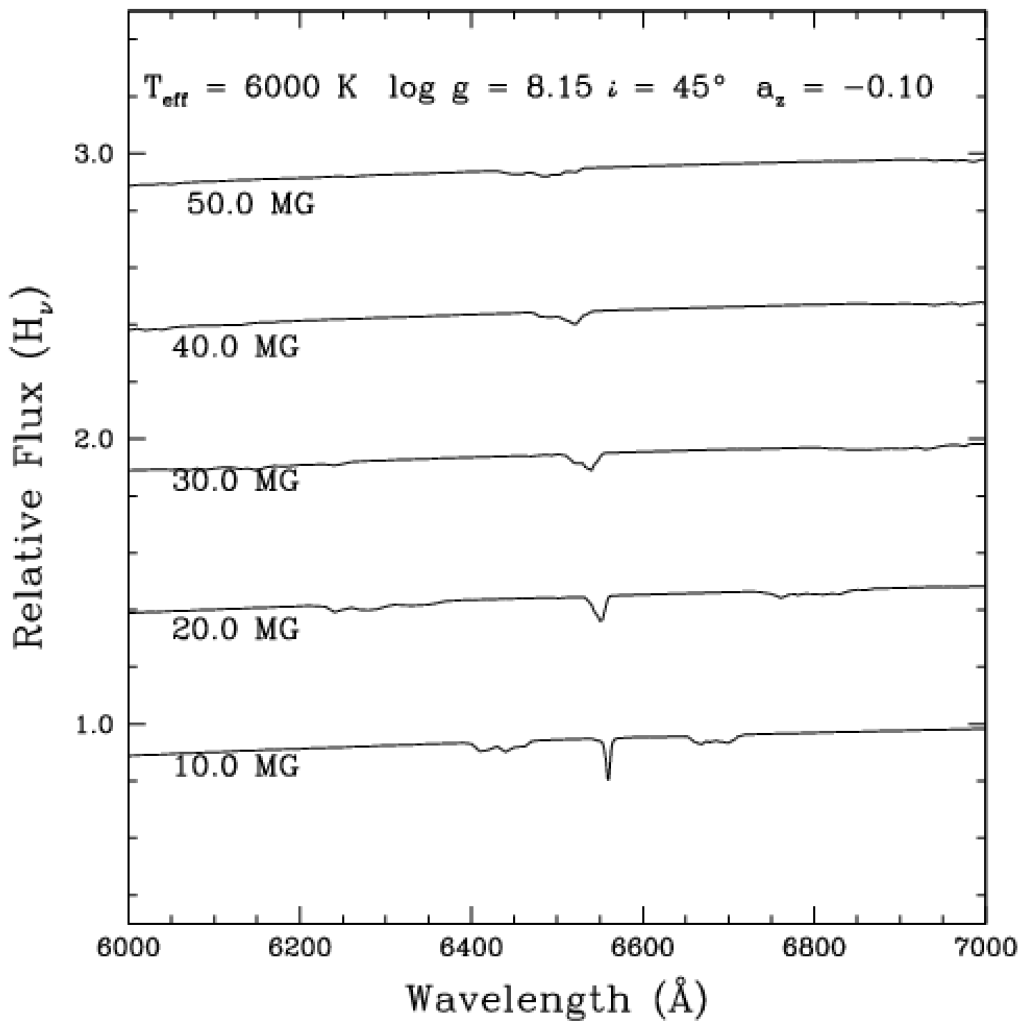}
    \caption{Synthetic spectra for a 0.75 $M_{\odot}$ WD and fixed magnetic field geometry. The left, middle, and right panels show $T_{\rm eff}=10,000, 8,000, 6,000$ K respectively. Each spectrum is labeled with its field strength, with weaker fields at the bottom of each panel.}
    \label{synthetic}
\end{figure*}

To better understand what field strengths we are not sensitive to, we plot synthetic spectra convolved with a 4 \AA\ width Gaussian profile. We set $\log{g} = 8.15$ and implement a typical geometry with $i = 45\degree$ and $a_z = -0.1$. We vary the field strength, checking low field strengths at high temperatures and high field strengths at low temperatures. Figure \ref{synthetic} shows our synthetic spectra at various $T_{\rm eff}$ and $B_d$. At $T_{\rm eff}=10,000$ K, it is very unlikely that we could detect field strengths below $\sim$0.5 MG. But around 0.5 MG, the line core begins to flatten, even though the individual $\sigma$ components would not be resolved (see for example the spectrum and fit for J1139+2613). At $T_{\rm eff}=8,000$ K, we can reliably detect field strengths up to $\sim$100 MG, as the H$\alpha$ feature has not completely vanished assuming our geometry. At $T_{\rm eff}=6,000$ K, both the $\pi$ and $\sigma$ components become very weak above 40 MG. Below this threshold, it is possible we can still detect fields based on the depth and the shape of the central line, however the absence of the split components would likely lead to degeneracies in the model fits. Note that these field strengths are not firm limits as the synthetic spectra do not incorporate observational noise, which would further increase the difficulty of detecting fields. The typical WD spectrum in the 100 pc SDSS sample has S/N $\geq$ 10. Hence any absorption lines shallower than $\sim$10\% of the continuum could be washed out by noise, further biasing our sample against fields in older objects.

It is worth noting that of the three MWDs in \citet{Bagnulo22} that are less massive than 0.75 $M_{\odot}$ and younger than 0.6 Gyr, two are DAH+dM systems and one is a ``DAH?'' that requires additional observations. High-field MWDs with M dwarf companions are exceedingly rare \citep{Liebert05,Liebert15}, while the two from the \citet{Bagnulo22} sample have field strengths of 0.15 and 0.01 MG, which are naturally below our detection limits. In addition, \citet{Kilic20} and \citet{Kilic25} made a specific cut in the color-magnitude diagram to remove white dwarfs with main-sequence companions. Hence it is not obvious that we are missing MWDs that have average masses and very young ages among the single WD population, which are the focus of this study.

\subsection{The Young, Massive, Strong-Field WDs}

The GMM implies there is a population of young, high field, massive MWDs but provides no physical constraint on their origin. It is unlikely that every single object in this population is a merger remnant, but rather mergers are the dominant formation channel for this population. Hence high mass single-star evolution could in principle produce some of these objects \citep{Richer19,Caiazzo20}. 

\citet{Camisassa24} showed that a main-sequence dynamo could generate a field that would reach the surface of the WD within $\sim$1 Gyr of cooling in ultramassive WDs. However there is currently no clear picture from an observational standpoint as to why the fields in these higher mass WDs are clearly stronger than those in more average mass WDs. \citet{Wick05} calculated mass distributions and space densities using different assumptions about the IFMR, and found that there are too few strongly-magnetic Ap and Bp main-sequence stars to explain the number of high-field MWDs ($B >$ 1 MG). Instead, they propose that if 40\% of main-sequence stars that are more massive than $\sim$4.5$M_{\odot}$ become high-field MWDs, this could reproduce the mass distribution and fraction of high-field MWDs. However, this would require that these stars have field strengths from 10$-$100 G. \citet{Bagnulo06} conducted a spectropolarimetric survey on 235 A and B stars, of which 97 are Ap and Bp stars. For the remaining 138 normal stars, none had detectable magnetic fields, though the median error in these measurements was 136 G. \citet{Shrolin02} similarly conducted a spectropolarimetric survey and found no field detection in any of their 22 normal B, A, and F stars. \citet{Fossati15} searched for fields in 50 O and B stars and only detected fields in 3 of them. Hence there does not appear to be a sufficient amount of magnetic stars that could explain the majority of our high mass, high field population, at least based on detectable surface fields in stars. Additionally, there would need to be an explanation as to why many ultramassive WDs are not magnetic. Only $\sim$32\% of the ultramassive sample from \citet{Jewett24} is magnetic, so some factor would need to generate a dynamo in certain high mass main-sequence stars but not in others. 

Theoretical modeling however does show some evidence of a relation between the mass of the star and the field strength buried within the stellar core. \citet{Brun05} conducted magnetohydrodynamic simulations on the inner 30\% of a 2$M_{\odot}$ main-sequence star, and found that strong convection is able to generate a field on the order of $\sim$60 kG. Similarly, \citet{Hidalgo24} performed simulations on the inner 20\% of a 2.2$M_{\odot}$ and also obtained field strengths of $\sim$60 kG.  \citet{Augustson16} meanwhile obtained field strengths exceeding 1 MG based on simulations of the inner 23\% of a 10$M_{\odot}$ star. Future modeling on a wide range of stellar masses as well as core rotation rates will be valuable for establishing a strong connection between the field generated in the core and the final WD field. 

Since we cannot measure the rotation periods in many of these objects, the tangential velocities could potentially identify merger products. However as we show in Figure \ref{fig16}, very few of the young, high mass objects have large tangential velocities. This is similar to the findings of \citet{Jewett24}, who detected large velocities in only five of their magnetic objects, and only 15\% of the total sample. Hence it does not appear that tangential velocities by themselves can isolate merger products, particularly in samples dominated by WDs in the thin disk such as the 100 pc sample.  

\citet{Bagnulo24} found that young, massive MWDs show no preference toward symmetric or non-symmetric fields about their rotation axes. This could be due to a number of variables, such as contributions from dynamos generated through mergers and single-star evolution, the initial geometry of the individual WDs during a merger, and the inability to measure extremely fast rotation periods with long spectropolarimetric cadences. They also found a clear lack of variability among the old, average mass targets with fields stronger than 10 MG. We have rotation period estimates for 12 of 27 (44\%) objects in our sample that are more massive than 1$M_{\odot}$, but only 13 of 135 (9.6\%) for everything below this mass. 82 of our 116  MWDs that do not have confirmed periods are less massive than $1M_{\odot}$ and have fields weaker than 10 MG. So at least through photometry, either the high mass WDs tend to be more variable than lower mass objects, or the lower mass WDs have longer periods that are simply harder to detect.

\subsection{The Old, Average Mass, Weak-Field WDs}

Our second population of MWDs is more interesting as the exact dynamo origin is unclear beyond the fact that these objects likely come from single-star evolution. The first question to answer is if the crystallization process can generate a magnetic field. Multiple theoretical studies have been conducted to address this, primarily by investigating the ability of convection during C/O phase separation to generate velocities capable of generating a dynamo. \citet{Montgomery24} found that the resulting velocities are lower than those found by \citet{Isern17} and \citet{Ginzburg22}, and thus could not form a dynamo. This could be resolved by fast rotation rates \citep{Schreiber21}, however \citet{Bagnulo24} found that these average-mass crystallized MWDs typically do not show variable fields, which makes it difficult (if not impossible) to determine the rotation periods of these objects. Other studies have shown that a dynamo could be generated but only at the onset of crystallization when efficient convection is present and before thermohaline convection becomes dominant \citep{Castro24,Fuentes23,Fuentes24}.

As shown in Figure \ref{fig7}, while many of our magnetic targets have begun core-crystallization, a non-negligible fraction has not. Of our 163 MWDs, the models suggest that about 47 (or 29\%) have not begun crystallization, 40 of which are less massive than 1$M_{\odot}$. Hence even if crystallization can spin up a dynamo, this mechanism alone cannot explain the source of magnetism in all of our lower mass objects. We assume an equal fraction of C and O in the core, which impacts the $T_{\rm eff}$ that WDs begin crystallizing. Assuming a higher O abundance leads to earlier crystallization times due to a larger coupling constant and average ion charge. However \citet{Obrien24} found that several MWDs in the 40 pc sample have not begun crystallizing despite using an O fraction of 0.8. Similarly, \citet{Hernandez24} found multiple MWDs with long rotation periods and typical masses that had not begun crystallizing. Hence it is unlikely that changing the oxygen abundance will allow a crystallization dynamo to explain all of our low mass MWDs, even if a crystallization-driven dynamo is plausible. 

A second issue arises with regards to when the magnetic field from a crystallization dynamo reaches the surface. The diffusion of the dynamo from the convective boundary to the surface is expected to take several Gyr \citep{Isern17,Ginzburg22}. \citet{Blatman24} calculated this diffusion time as a function of mass and compared this result to the cooling ages of MWDs within the volume-limited sample from \citet{Bagnulo22}. While several MWDs that have fields strengths $>$ 1 MG first appear after the breakout time, many MWDs still appear before the breakout should happen. They also assumed that the field starts to diffuse right at the crystallization onset from a convective boundary close to the surface. \citet{Castro24} on the other hand found that the convective boundary is located deeper within the WD, leading to significantly longer breakout times. Hence the calculation from \citet{Blatman24} is actually a ``best-case'' scenario. So it seems that when accounting for the breakout time, crystallization by itself is insufficient. 

On the other hand, core-convective fields that are trapped within main-sequence stars should reach the surface faster than a crystallization-driven dynamo by several Gyr depending on the mass \citep{Camisassa24}. While uncertainties in the input physics results in large uncertainties in the emergence time, this provides a promising alternative to the crystallization dynamo which appears to fail for at least some of the low mass MWDs. \citet{Camisassa24}  compared their predicted emergence timescales to the volume-limited sample of MWDs \citep{Bagnulo22} and found that a significantly higher fraction could be explained by this core-convective dynamo rather than the crystallization dynamo.

\begin{figure*}
    \centering
    \includegraphics[width=7in,clip=true]{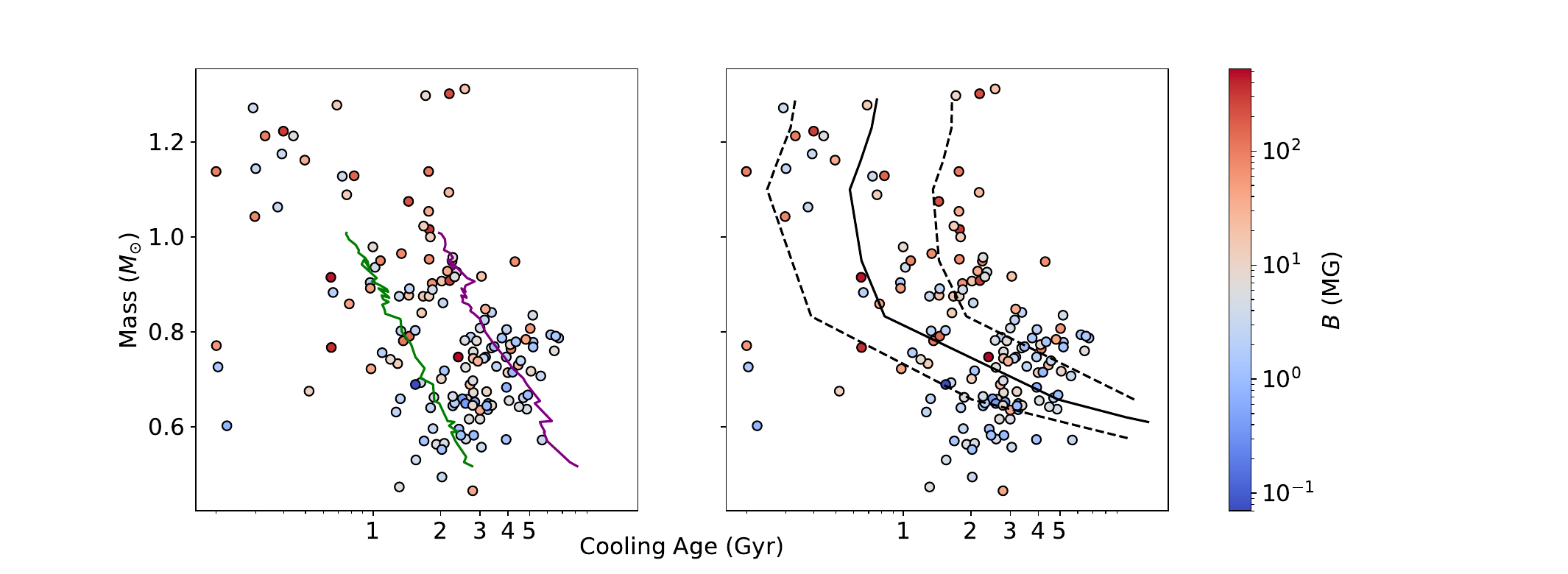}
    \caption{Mass as a function of a cooling age with points colored by field strength. The left panel shows the onset of crystallization (green line) and the breakout time for a crystallization-induced dynamo (purple line) as calculated by \citep{Blatman24}. The right panel shows the emergence of a core-convection generated dynamo produced on the main sequence (black solid line) as calculated by \citet{Camisassa24}. The dashed lines represent an error of $\pm0.3$ dex. The majority of our low-mass objects are not old enough for a hypothetical crystallization dynamo to have reached the surface. The main sequence dynamo on the other hand almost splits the low-mass population down the middle, with many objects within the errors. Hence it can potentially explain the occurrence of magnetism for more of our low-mass targets.}
    \label{fig21}
\end{figure*}

In Figure \ref{fig21} we plot the mass-age distribution with points colored by field strength, along with the emergence times expected from a crystallization and core-convective dynamo found by \citet{Blatman24} and \citet{Camisassa24}  respectively. It is clear from these plots that the core-convective dynamo can explain at least some of the MWDs that crystallization cannot. Only 21 of our 101 MWDs with field estimates and that are less than 0.85$M_{\odot}$ lie to the right of the crystallization dynamo breakout line, while an additional 22 lie to the right of the core-convective dynamo emergence line. Many of the objects below $\sim$0.7$M_{\odot}$ and younger than $\sim$4 Gyr could also be explained by the core-convective dynamo given the uncertainties found by \citet{Camisassa24}. This mechanism could also explain some of the high mass MWDs if they in fact do not come from mergers.

\section{Conclusion}
We have performed a model atmosphere analysis on the MWDs in the SDSS 100 pc sample \citep{Kilic25}. Our sample contains 163 MWDs, 87 of which are new discoveries. We have magnetic field estimates either from our model fits or from the literature for 152 of the 163 MWDs. This analysis is thus the largest volume-limited survey of MWDs to date. To explore the two formation channel hypothesis in our sample, we split our sample into two groups based on cooling age: one that contains targets younger than 2 Gyr, and one for targets older than 2 Gyrs. We find statistically significant evidence of two separate populations of MWDs through the use of K-S tests and a GMM. Our K-S tests provide low p-values ($< 0.05$) when comparing the field strengths of each age group provided we use an age cutoff between 1.0 and 3.5 Gyr. Our GMM analysis finds two populations centered at 2.9 Gyr, 0.71 $M_{\odot}$, 3.7 MG and 1.8 Gyr, 0.96 $M_{\odot}$, 84 MG respectively.

These populations show clear differences in their physical properties because they are dominated by MWDs that come from two distinct formation channels. The young, high mass, high field strength population is dominated by merger remnants, while the old, average mass, weak field strength population is dominated by targets formed through single-star evolution. Although there is little overlap at the 1$\sigma$ level in the populations based on our GMM analysis, we cannot completely rule out single-star evolution products in our high mass population or binary evolution products in the low mass population. The presence of two populations was proposed by \citet{Valyavin99} and confirmed by \citet{Bagnulo22}. This work provides further statistically significant evidence of these distinct populations with the use of a large and robust sample.

In addition to the masses, cooling ages, and field strengths of our MWDs, we use rotation periods and tangential velocities to provide further insight into the formation channel. We have rotation periods for 26 targets using data from TESS, ZTF, and values reported in the literature, 23 of which are shorter than 6 hrs and 15 have masses $> 0.9M_{\odot}$. These high mass, fast-spinning objects are thus extremely likely merger products. However, only 4 of these 15 have tangential velocities $>$ 50 km/s, and the sample shows a typical increase in velocity dispersion for older targets. Hence we are unable to use kinematics to identify merger remnants for this thin disk dominated sample. 

While binary evolution can likely explain the majority of the young population, some ambiguity remains for the old population. These objects likely come from single-star evolution, but whether these dynamos come from crystallization or from earlier stellar evolution stages is still up for debate. However, when accounting for breakout times, a noticeably larger fraction of our older targets can be explained through a dynamo that forms on the main-sequence rather than crystallization. This result is in strong agreement with the findings of \citet{Blatman24} and \citet{Camisassa24}. Uncertainties in the treatment of convective boundaries and the IFMR leads to large uncertainties in the breakout times for this core-convective dynamo. Nevertheless, crystallization itself seems unlikely to generate the magnetic fields we see in many of our low mass MWDs given the long breakout times for this dynamo.

The primary bias in our survey is against objects with fields too weak to produce clear Zeeman-splitting and WDs that do not show absorption features in their spectra. The latter group consists of He-dominated atmosphere WDs cooler than $\sim$11,000 K, and H-dominated atmosphere WDs cooler than $\sim$5000 K. If we expect to see absorption features based on the atmosphere composition and $T_{\rm eff}$ but do not, we can infer the presence of a strong field. Otherwise, these WDs with featureless spectra cannot be confirmed as magnetic without spectropolarimetry. Hence we are biased against older WDs with field strengths greater than 100 MG, and all MWDs with field strengths weaker than $\sim$1 MG. While the biases in our sample prevent us from obtaining a complete view of specific features within the MWD population, such as the evolution of the incidence and strength of fields among 
lower-mass MWDs, the statistical analysis shows a clear distinction between the physical parameters of MWDs. The two distinct categories of MWDs are not just two blobs in the age-mass diagram, but two categories of stars both of which have their own origin and evolutionary pattern. The most important reason for the distinction between hot and cooler objects is that low mass WDs frequently become strongly magnetic with age.

Current and upcoming surveys have potential to discover new MWDs and significantly improve the statistics that we have explored in this work. DESI DR1 provided optical spectra for $\sim$50,000 WDs from the \citet{Gentile19} catalog down to $G \simeq 20$ mag \citep{desi}. Similarly, SDSS-V will target nearly 200,000 WDs \citep{sdssv} while the 4MOST survey \citep{4Most} will obtain spectra for thousands of WDs in the Southern hemisphere. Depending on their brightness, many of these objects can then be followed up with spectropolarimetry to detect fields that cannot be found through conventional Zeeman-splitting. After many years of ambiguity as to the source of magnetism in WDs, a clearer picture is starting to emerge thanks to both observational surveys and theoretical dynamo simulations. With upcoming large surveys and continued improvement in the treatment of internal physics, the future of understanding magnetism in WDs is highly promising. \\

%\begin{acknowledgments}
We thank the anonymous referee for their feedback that significantly improved this article. We thank J. Rupert for obtaining the MDM data as part of the OSMOS queue. This work is supported in part by the NSF under grant AST-2205736, the NASA under grants 80NSSC22K0479, 80NSSC24K0380, and 80NSSC24K0436, the NSERC Canada, the Fund FRQ- NT (Québec), and the Smithsonian Institution. This work is supported by grant RYC2021-032721-I, funded by MCIN/AEI/10.13039/501100011033 and by the European Union NextGenerationEU/PRTR, the AGAUR/Generalitat de Catalunya grant SGR-386/2021 and by the Spanish MINECO grant PID2023-148661NB-I00.

The Apache Point Observatory 3.5-meter telescope is owned and operated by the Astrophysical Research Consortium.

Based on observations obtained at the MMT Observatory, a joint facility of the Smithsonian Institution and the University of Arizona.
This paper includes data gathered with the 6.5 meter Magellan Telescopes located at Las Campanas Observatory, Chile.

This work is based on observations obtained at the MDM Observatory, operated by Dartmouth College, Columbia University, Ohio State University, Ohio University, and the University of Michigan. The authors are honored to be permitted to conduct astronomical research on Iolkam Du’ag (Kitt Peak), a mountain with particular significance to the Tohono O’odham.

Based on observations obtained at the international Gemini Observatory, a program of NSF’s NOIRLab, which is managed by the Association of Universities for Research in Astronomy (AURA) under a cooperative agreement with the National Science Foundation on behalf of the Gemini Observatory partnership: the National Science Foundation (United States), National Research Council (Canada), Agencia Nacional de Investigación y Desarrollo (Chile), Ministerio de Ciencia, Tecnología e Innovación (Argentina), Ministério da Ciência, Tecnologia, Inovaçōes e Comunicaçōes (Brazil), and Korea Astronomy and Space Science Institute (Republic of Korea).

Facilities: ARC 3.5m (KOSMOS spectrograph), FLWO 1.5m (FAST spectrograph), Gemini (GMOS spectrograph), Magellan:Baade (MagE spectrograph), MDM (OSMOS), MMT (Blue Channel spectrograph)

%\end{acknowledgments}

\newpage

\bibliographystyle{aasjournal}
\bibliography{Moss_Mag100pc.bib}

\newpage
\appendix

\startlongtable
\centering

%{\setlength{\tabcolsep}{2pt}
%\begin{longrotatetable}
\begin{deluxetable}{cccccccccccc}
\tabletypesize{\scriptsize}
\tablecaption{\label{tab3} Physical parameters for our MWD sample.}
\tablecolumns{12} \tablewidth{0pt}
\centering
\tablehead{\colhead{SDSS ID} & \colhead{Gaia DR3 ID} & \colhead{Composition} & \colhead{Spectral Type} & \colhead{$T_{\rm eff}$} & \colhead{Mass} & \colhead{Cooling Age} & \colhead{B} & \colhead{$i$} & \colhead{$a_z$} & \colhead{Source} & \colhead{Note} \\
\colhead{} & \colhead{} & \colhead{} & \colhead{} & \colhead{(K)} & \colhead{($M_{\odot}$)} & \colhead{(Gyr)} & \colhead{(MG)} & \colhead{(\degree)} & \colhead{($R/R_*$)} & \colhead{} & \colhead{}}
\startdata
J000603.80+075514.4 & 2746037712074342784 & H & DAH & $8607\pm75$ & $0.733\pm0.018$ & $1.288\pm0.050$ & 13.7 & 48 & 0.30 & 7 & \\
J000637.93+310415.7 & 2861452348130844160 & H & DAH? & $25442\pm522$ & $1.138\pm0.010$ & $0.200\pm0.014$ & 93.0 & 0 & 0.50 & 1 & X\\            
J000728.91+340341.4 & 2875903332533220992 & H & DAP & $5917\pm15$ & $0.807\pm0.005$ & $5.026\pm0.047$ & 62.0 & 0 & -0.29 & 1 & X\\            
J001328.72+074844.6 & 2751750087297310592 & H & DAH & $6219\pm31$ & $0.672\pm0.012$ & $2.801\pm0.124$ & 8.0 & 45 & -0.13 & 1 & \\            
J002128.60+150223.8 & 2792315366213367296 & H & DAH & $7119\pm39$ & $0.747\pm0.012$ & $2.399\pm0.102$ & 527.3 & 28 & 0.16 & 10 & \\            
J005045.81-032655.6 & 2529337507976700928 & H & DAH? & $23916\pm355$ & $1.213\pm0.006$ & $0.330\pm0.017$ & 99.0 & 15 & 0.23 & 1 & X\\           
J005233.73+150916.5 & 2778069062775645184 & H & DAH & $5742\pm50$ & $0.636\pm0.032$ & $3.253\pm0.371$ & 1.0 & 0 & 0.23 & 1 & X\\
J010624.44+150626.4 & 2783309850585780352 & H & DAH: & $5758\pm57$ & $0.778\pm0.033$ & $5.178\pm0.30$ & 1.6 & 60 & 0.03 & 1 & X\\            
J015013.17+283556.2 & 299265624604662656 & H & DAH & $12046\pm101$ & $0.979\pm0.008$ & $0.999\pm0.035$ & 8.0 & 60 & 0.30 & 9 & \\            
J015342.00+180857.9 & 91980130553425536 & H & DAH & $9147\pm49$ & $0.875\pm0.007$ & $1.674\pm0.051$ & 19.0 & 45 & 0.49 & 1 & \\            
J015453.71+253739.3 & 297399444134124032 & H & DAH & $6778\pm53$ & $0.725\pm0.019$ & $2.587\pm0.177$ & 6.0 & 90 & 0.48 & 1 & \\
J021148.21+211548.2 & 99498964725981440 & H & DAH & $12244\pm135$ & $1.075\pm0.007$ & $1.440\pm0.054$ & 207 & 45 & 0.4 & 9 & \\
J021407.27-050548.1 & 2488226252817987584 & H & DAH & $9005\pm71$ & $0.907\pm0.010$ & $2.024\pm0.077$ & 19.0 & 15 & 0.47 & 1 & \\
J022742.58+442444.1 & 341432690100603904 & H & DAH: & $14444\pm107$ & $0.602\pm0.006$ & $0.223\pm0.006$ & 0.8 & 90 & -0.04 & 1 & X\\
J025506.14-071551.4 & 5180494685197598464 & H & DAH & $7149\pm55$ & $0.718\pm0.024$ & $2.080\pm0.154$ & 1.4 & 7 & -0.28 & 2 & \\
J025930.85+034058.6 & 2021898804878976 & H & DAH: & $5695\pm29$ & $0.574\pm0.015$ & $2.596\pm0.089$ & 4.1 & 60 & -0.07 & 1 & X\\
J030350.63+060749.0 & 6963383233077632 & H & DAH? & $13150\pm542$ & $0.915\pm0.020$ & $0.649\pm0.077$ & 445 & 90 & -0.45 & 1 & X\\
J031613.91-081637.5 & 5167510797199003904 & H & DAH: & $6482\pm34$ & $0.805\pm0.008$ & $3.944\pm0.076$ & 2.3 & 90 & 0.41 & 1 & X\\
J032619.23+133108.6 & 17709047809907584 & H & DAH & $11351\pm101$ & $0.936\pm0.010$ & $1.021\pm0.037$ & 4.0 & 15 & 0.30 & 9 & \\
J035324.54+103728.0 & 3303137301563100544 & H & DAH? & $5922\pm47$ & $0.784\pm0.029$ & $4.802\pm0.230$ & 41.0 & 75 & -0.09 & 1 & X\\
J035734.99+191549.3 & 50456318016101120 & H & DAH & $5905\pm35$ & $0.616\pm0.014$ & $2.684\pm0.114$ & 6.4 & 0 & -0.22 & 1 & \\
J042229.75+165706.5 & 3313861186362477824 & H & DAH & $6623\pm39$ & $0.530\pm0.013$ & $1.553\pm0.042$ & 4.0 & 60 & 0.42 & 1 & \\
J052054.17+011442.2 & 3222138444609365248 & H & DAH & $6590\pm55$ & $0.766\pm0.019$ & $3.371\pm0.172$ & 6.0 & 75 & 0.43 & 1 & \\
J055119.47-001021.4 & 3218697767783768320 & log(C/He) = -6.07 & DQH & $6228\pm4$ & $0.674\pm0.001$ & $3.210\pm0.012$ & 10 & NA & NA & 4 & \\
J060755.76+341526.1 & 3452373568124842752 & H & DAH & $21139\pm340$ & $1.175\pm0.006$ & $0.392\pm0.021$ & 3.2 & 45 & 0.3 & 9 & \\
J064400.57+092605.1 & 3326650224581677312 & H & DAH & $6217\pm26$ & $0.658\pm0.007$ & $2.627\pm0.065$ & 6.0 & 60 & -0.11 & 1 & \\
J065757.54+660505.7 & 1102382143399269632 & H & DAH: & $5183\pm29$ & $0.465\pm0.016$ & $2.784\pm0.124$ & 41.0 & 60 & -0.25 & 1 & X\\
J071816.41+373139.1 & 898348313253395968 & H & DAH? & $33942\pm1411$ & $1.317\pm0.007$ & $0.203\pm0.029$ & NA & NA & NA & 9 & \\
J074213.37+315703.4 & 880354496226790400 & H & DAH & $9426\pm60$ & $0.722\pm0.010$ & $0.979\pm0.023$ & 41.4 & 11 & -0.16 & 2 & \\
J074402.82+154442.7 & 3165713981772896000 & H & DAH: & $5270\pm30$ & $0.573\pm0.021$ & $3.924\pm0.368$ & 1.2 & 15 & -0.35 & 1 & X\\
J074842.48+112502.0 & 3151443115414582016 & log(H/He) = -4.20 & DCP & $8983\pm57$ & $0.781\pm0.009$ & $1.364\pm0.034$ & 112.0 & 45 & 0.10 & 1 & X\\
J075636.88-004525.4 & 3083915298828722304 & H & DAH: & $5966\pm32$ & $0.595\pm0.011$ & $2.417\pm0.069$ & 1.0 & 15 & 0.06 & 1 & X\\
J080042.46+065542.1 & 3144237908341731712 & H & DAH & $10623\pm112$ & $1.016\pm0.012$ & $1.782\pm0.077$ & 317.0 & 75 & 0.23 & 1 & \\
J080323.83+163356.9 & 655245124121246720 & H & DAH & $5456\pm33$ & $0.661\pm0.017$ & $4.682\pm0.264$ & 2.6 & 75 & 0.33 & 1 & \\
J080359.94+122944.0 & 653051877600013312 & H & DAH & $19168\pm332$ & $1.162\pm0.008$ & $0.496\pm0.030$ & 35.0 & 15 & 0.30 & 9 & \\
J080643.94+405837.7 & 921642532239821440 & log(Ca/He) = -9.16,  & DZH & $6091\pm39$ & $0.649\pm0.022$ & $3.263\pm0.0220$ & 0.8 & NA & NA & 8 & \\
 & & log(He/H) = 3.0 & & & & & & & & & \\
J081802.74+414410.2 & 915672081021495552 & H & DAH & $8659\pm59$ & $0.791\pm0.015$ & $1.454\pm0.052$ & 143.0 & 30 & 0.09 & 1 & X\\
J083531.17+533230.9 & 1030289712881895296 & H & DAH & $8818\pm148$ & $0.840\pm0.022$ & $1.648\pm0.148$ & 14.4 & 10 & -0.46 & 2 & \\
J084155.74+022350.6 & 3079104660578372480 & H & DAH & $6905\pm74$ & $0.758\pm0.025$ & $2.799\pm0.230$ & 6.5 & 51 & 0.30 & 7 & \\
J084214.99-022226.7 & 3072348715677121280 & H & DAH? & $23910\pm212$ & $1.162\pm0.004$ & $0.263\pm0.007$ & NA & NA & NA & 9 & \\
J084452.43+103537.5 & 598769843231677056 & H & DAH: & $6313\pm31$ & $0.659\pm0.017$ & $2.503\pm0.133$ & 0.4 & 15 & 0.22 & 1 & X\\
J085106.12+120157.8 & 604972428842238080 & H & DAH & $11209\pm83$ & $0.904\pm0.008$ & $0.971\pm0.025$ & 1.8 & 60 & 0.3 & 9 & \\
J085618.94+161103.6 & 611401999180118528 & log(H/He) = -5.50 & DBAH & $16984\pm497$ & $0.726\pm0.026$ & $0.204\pm0.022$ & 1.50 & 0 & -0.40 & 7 & \\
J085649.67+253441.1 & 691604221304319744 & H & DAH & $11325\pm118$ & $0.950\pm0.012$ & $1.079\pm0.055$ & 85.0 & 19 & -0.20 & 2 & \\
J085830.85+412635.0 & 912718071240545152 & H & DAH & $7097\pm28$ & $0.841\pm0.006$ & $3.379\pm0.057$ & 2.6 & 41 & 0.3 & 7 & \\
J092213.46+050438.3 & 585488567401589248 & H & DAH & $5760\pm24$ & $0.683\pm0.010$ & $3.936\pm0.121$ & 0.7 & 30 & 0.19 & 1 & X\\
J092433.23+052106.8 & 585513959248023936 & H & DAH: & $6185\pm40$ & $0.649\pm0.020$ & $2.587\pm0.161$ & 0.4 & 15 & 0.23 & 1 & X\\
J094122.18+071617.9 & 3854183431001210112 & H & DAH: & $5917\pm23$ & $0.714\pm0.009$ & $3.992\pm0.105$ & 13.4 & 0 & -0.24 & 1 & X\\
J100759.81+162349.6 & 621979502901486720 & H & DAH & $10392\pm83$ & $1.000\pm0.011$ & $1.801\pm0.067$ & 13.6 & 42 & 0.30 & 7 & \\
J101131.94+372713.1 & 801851053926004608 & H & DAH: & $6398\pm35$ & $0.644\pm0.019$ & $2.269\pm0.104$ & 2.0 & 30 & 0.49 & 1 & X\\
J101145.20+495834.3 & 823804585985883776 & H & DAH & $6155\pm30$ & $0.747\pm0.014$ & $3.929\pm0.123$ & 2.0 & 60 & -0.23 & 1 & \\
J101529.62+090703.7 & 3876618892751168000 & H & DAH & $7146\pm63$ & $0.808\pm0.017$ & $3.002\pm0.150$ & 4.5 & 47 & 0.30 & 7 & \\
J101631.72+181754.7 & 624090737025312000 & H & DAH? & $7310\pm80$ & $0.848\pm0.024$ & $3.171\pm0.183$ & 34.0 & 30 & 0.01 & 1 & X\\
J101805.04+011123.6 & 3832329434808415744 & H & DAH & $10869\pm77$ & $0.965\pm0.007$ & $1.339\pm0.047$ & 76.1 & 16 & 0.05 & 7 & \\
J102054.11+362647.1 & 753840603821390848 & H & DAH & $17361\pm187$ & $0.771\pm0.008$ & $0.200\pm0.008$ & 59.8 & 69 & 0.15 & 7 & \\
J102837.65+145101.8 & 3888929097950924416 & H & DAH & $5836\pm30$ & $0.582\pm0.012$ & $2.463\pm0.063$ & 1.2 & 60 & 0.00 & 1 & X\\
J102907.55+112718.3 & 3882611201058534400 & H & DAH & $6796\pm28$ & $0.744\pm0.005$ & $2.798\pm0.061$ & 19.2 & 46 & 0.30 & 7 & \\
J103430.16+032736.4 & 3856950175919062144 & H & DAH & $15110\pm167$ & $1.089\pm0.006$ & $0.763\pm0.033$ & 13.0 & 75 & 0.30 & 9 & \\
J103532.53+212603.6 & 721244821406496640 & H & DAH & $6949\pm46$ & $0.640\pm0.012$ & $1.806\pm0.049$ & 2.6 & 48 & 0.30 & 7 & \\
J103941.53-032534.2 & 3778305545157841152 & H & DAH? & $19792\pm235$ & $1.043\pm0.008$ & $0.297\pm0.011$ & 86.0 & 15 & 0.49 & 1 & X\\
J105320.18-232926.0 & 3549053213238728704 & H & DAH: & $5344\pm33$ & $0.637\pm0.02$ & $4.853\pm0.338$ & 4.3 & 15 & -0.17 & 1 & X\\
J105544.90+211105.0 & 3988212592756945152 & H & DAH & $6051\pm24$ & $0.652\pm0.008$ & $2.845\pm0.079$ & 0.5 & 7 & -0.25 & 2 & \\
J105556.77+252556.1 & 729726763340729600 & H & DAH & $7069\pm35$ & $0.701\pm0.016$ & $2.017\pm0.087$ & 11.0 & 30 & 0.50 & 1 & X\\
J105607.14+385257.6 & 775531009004593664 & H & DAH: & $5820\pm30$ & $0.730\pm0.012$ & $4.443\pm0.137$ & 25.0 & 45 & -0.16 & 1 & \\
J105709.82+041130.4 & 3815759279181364864 & H & DAH & $7870\pm43$ & $0.631\pm0.010$ & $1.267\pm0.028$ & 2.5 & 41 & 0.38 & 10 & \\
J110505.75+522522.2 & 842469143761181824 & H & DAH? & $17588\pm140$ & $1.154\pm0.005$ & $0.633\pm0.023$ & NA & NA & NA & 9 & \\
J111638.37+311921.2 & 756823437133003136 & H & DAH & $5539\pm28$ & $0.794\pm0.011$ & $6.197\pm0.166$ & 1.2 & 0 & 0.05 & 1 & X \\
J112328.49+095619.4 & 3915026861134449664 & H & DAH & $9323\pm80$ & $0.756\pm0.016$ & $1.098\pm0.040$ & 1.9 & 70 & 0.30 & 7 & \\
J112605.90+090628.9 & 3914160793864337280 & H & DAH & $10639\pm112$ & $1.094\pm0.011$ & $2.179\pm0.065$ & 26.1 & 2 & 0.28 & 7 & \\
J111341.32+014641.7 & 3810552434493888768 & log(C/He) = -6.50 & DQH & $5226\pm16$ & $0.866\pm0.009$ & $7.314\pm0.032$ & NA & NA & NA & 1 & \\
J113756.51+574022.4 & 845982667526493440 & H & DAH & $7586\pm38$ & $0.693\pm0.008$ & $1.633\pm0.031$ & 2.8 & 0 & 0.17 & 10 & \\
J113926.44+261320.5 & 4017476365313408512 & H & DAH: & $6570\pm35$ & $0.747\pm0.010$ & $3.176\pm0.100$ & 0.4 & 0 & 0.25 & 1 & X\\
J114207.26+662443.3 & 1057020382445488512 & H & DAH: & $5435\pm54$ & $0.667\pm0.032$ & $4.902\pm0.489$ & 1.0 & 45 & 0.22 & 1 & X\\
J114439.54+662928.5 & 1056998259069523584 & H & DAH & $6717\pm30$ & $0.473\pm0.007$ & $1.310\pm0.021$ & 6.0 & 75 & 0.50 & 1 & \\
J120125.41+084800.5 & 3900013579612540672 & H & DAH & $7207\pm64$ & $0.789\pm0.024$ & $2.728\pm0.198$ & 2.2 & 12 & -0.1 & 2 & \\
J120327.49+420602.6 & 1537203704733382016 & H & DAH & $5773\pm22$ & $0.494\pm0.014$ & $2.030\pm0.055$ & 3.0 & 30 & -0.49 & 1 & X \\
J120332.77+591011.5 & 1576088036528753408 & H & DAH? & $9872\pm78$ & $0.953\pm0.011$ & $1.779\pm0.076$ & 73.0 & 15 & 0.50 & 1 & X\\
J121033.23+221402.6 & 4001466277717538432 & H & DAH & $12636\pm268$ & $0.883\pm0.010$ & $0.663\pm0.026$ & 2.0 & 86 & -0.30 & 7 & \\
J121456.38-023402.9 & 3694399755554510720 & log(Ca/He) = -8.26, & DZH & $5362\pm25$ & $0.787\pm0.010$ & $6.733\pm0.069$ & 2.12 & NA & NA & 8 & \\
 & & log(He/H) = 1.0 & & & & & & & & & \\
J121548.73-005637.5 & 3698040337339931776 & H & DAH & $6586\pm59$ & $0.744\pm0.025$ & $3.117\pm0.234$ & 3.0 & 90 & 0.00 & 1 & \\
J121735.25+082810.1 & 3902183809407583872 & H & DAH & $18654\pm268$ & $1.063\pm0.008$ & $0.376\pm0.017$ & 3.2 & 60 & 0.30 & 9 & \\
J122920.15+412611.5 & 1534719152051775616 & H & DAH & $6306\pm47$ & $0.563\pm0.025$ & $1.917\pm0.098$ & 6.0 & 45 & -0.49 & 1 & \\
J124004.84-100242.1 & 3578787203109091968 & H & DAH & $5721\pm21$ & $0.616\pm0.011$ & $3.000\pm0.123$ & 5.0 & 30 & -0.50 & 1 & X\\
J124146.05-150629.4 & 3526928874624208128 & H & DAH & $7920\pm105$ & $0.917\pm0.020$ & $3.047\pm0.146$ & 19.4 & 0 & 0.42 & 1 & X\\
J125446.22+491818.9 & 1555410036744248832 & H & DAH & $6547\pm69$ & $0.769\pm0.022$ & $3.473\pm0.207$ & 1.0 & 15 & 0.04 & 1 & \\
J125715.54+341439.4 & 1515693894335906688 & H & DAH & $9557\pm41$ & $0.877\pm0.006$ & $1.444\pm0.036$ & 19.7 & 72 & 0.29 & 7 & \\
J130033.46+590406.9 & 1578748824604827648 & H & DAH & $6132\pm35$ & $0.565\pm0.011$ & $2.079\pm0.051$ & 4.8 & 35 & -0.01 & 7 & \\
J131426.37+173228.0 & 3937174942327932544 & H & DAH & $7160\pm57$ & $0.825\pm0.017$ & $3.142\pm0.138$ & 2.6 & 14 & -0.35 & 2 & \\
J133250.73+011706.3 & 3711214067185666560 & H & DAH & $7021\pm27$ & $0.781\pm0.006$ & $2.896\pm0.058$ & 7.4 & 60 & -0.07 & 5 & \\
J133359.84+001655.0 & 3662951038644235776 & log(C/He) = -6.32 & DQH & $5180\pm17$ & $0.912\pm0.01$ & $7.409\pm0.029$ & NA & NA & NA & 1 & \\
J133431.97+602815.6 & 1662778377985657344 & H & DAH & $5853\pm51$ & $0.635\pm0.021$ & $3.001\pm0.225$ & 37.0 & 90 & -0.26 & 1 & \\
J135118.59+171943.9 & 1244753299874351488 & H & DAH & $9111\pm89$ & $0.949\pm0.012$ & $2.252\pm0.088$ & 103.0 & 60 & 0.37 & 1 & \\
J135141.13+541947.3 & 1561389253988468224 & H & DAH & $11423\pm78$ & $0.767\pm0.007$ & $0.652\pm0.014$ & 368.5 & 34 & 0.07 & 7 & \\
J140640.31+222456.2 & 1253540837322222336 & log(Ca/He) = -8.40,  & DZH & $5664\pm33$ & $0.665\pm0.017$ & $4.850\pm0.172$ & NA & NA & NA & 1 & \\
 & & log(He/H) = -30.0 & & & & & & & & & \\
J141737.99+180537.8 & 1233598174590839296 & H & DAH & $9053\pm85$ & $0.928\pm0.012$ & $2.147\pm0.088$ & 37.0 & 45 & 0.28 & 1 & \\
J141857.13+193530.3 & 1239963487922089600 & H & DAH? & $6147\pm35$ & $0.764\pm0.011$ & $4.124\pm0.111$ & 89.0 & 0 & -0.11 & 1 & X\\
J141906.19+254356.5 & 1258934014870979712 & H & DAH & $9062\pm52$ & $0.802\pm0.010$ & $1.331\pm0.033$ & 2.6 & 55 & 0.38 & 10 & \\
J141938.91+544313.3 & 1608747311444717440 & H & DAH: & $5978\pm64$ & $0.835\pm0.020$ & $5.159\pm0.188$ & 4.4 & 0 & -0.12 & 1 & X\\
J143019.05+281100.9 & 1280674894509973760 & H & DAH & $9163\pm77$ & $0.957\pm0.012$ & $2.268\pm0.080$ & 6.3 & 6 & 0.16 & 10 & \\
J143352.48+510058.2 & 1604200899583092992 & H & DAH & $5585\pm65$ & $0.655\pm0.029$ & $4.042\pm0.409$ & 5.4 & 60 & -0.41 & 1 & X\\
J144921.34+205421.9 & 1238352149340737536 & H & DAH: & $5723\pm50$ & $0.768\pm0.022$ & $5.177\pm0.235$ & 1.3 & 15 & -0.39 & 1 & X\\
J145335.93+422524.6 & 1489504004102020864 & H & DAH: & $5833\pm36$ & $0.715\pm0.015$ & $4.194\pm0.170$ & 0.9 & 30 & 0.04 & 1 & X\\
J150356.96+462721.1 & 1587150257375155584 & H & DAH & $6110\pm42$ & $0.552\pm0.016$ & $2.027\pm0.069$ & 1.2 & 0 & 0.15 & 1 & X\\
J150613.45+215051.5 & 1261705673230699008 & H & DAZH & $5614\pm38$ & $0.582\pm0.027$ & $2.812\pm0.206$ & 0.9 & 0 & 0.13 & 1 & X\\
J150942.07+141121.3 & 1183473535423719296 & H & DAH & $6403\pm22$ & $0.650\pm0.008$ & $2.313\pm0.048$ & 1.9 & 60 & -0.03 & 1 & \\
J151130.17+422023.0 & 1393328553789078784 & H & DAH & $10924\pm108$ & $1.023\pm0.010$ & $1.681\pm0.067$ & 14.0 & 40 & 0.30 & 7 & \\
J151625.07+280320.9 & 1271649969930799872 & H & DAH & $7944\pm47$ & $0.659\pm0.009$ & $1.324\pm0.030$ & 2.3 & 33 & 0.00 & 2 & \\
J153843.10+084238.3 & 1164767677244452096 & H & DAH & $8998\pm94$ & $0.875\pm0.017$ & $1.778\pm0.116$ & 11.5 & 49 & 0.29 & 7 & \\
J154020.79+042421.0 & 4427716454423219968 & H & DAH: & 5471 & 0.791 & 6.528 & 1.5 & 15 & -0.47 & 1 & X\\
J154315.10+302133.6 & 1273088783971336576 & H & DAH & $15661\pm198$ & $1.129\pm0.006$ & $0.824\pm0.043$ & 160 & 15 & 0.30 & 9 & \\
J154345.65-024741.2 & 4402794756366307328 & log(Ca/He) = -8.99, & DZH & $6585\pm34$ & $0.549\pm0.012$ & $1.796\pm0.041$ & NA & NA & NA & 1 & \\
 & &  log(He/H) = 30.0 & & & & & & & & & \\
J154855.05+245113.0 & 1219699145026398848 & H & DAH & $21662\pm386$ & $1.213\pm0.006$ & $0.441\pm0.028$ & 6.4 & 60 & 0.2 & 9 & \\
J155857.69+041704.9 & 4425632987265111680 & H & DAH & $6990\pm20$ & $0.948\pm0.005$ & $4.295\pm0.032$ & 69.8 & 10 & -0.16 & 2 & \\
J160437.35+490809.2 & 1400157173832960384 & H & DAH & $9215\pm79$ & $0.902\pm0.011$ & $1.835\pm0.083$ & 67.1 & 40 & 0.30 & 7 & \\
J160604.03+550925.5 & 1429284061486392576 & H & DAH & $5753\pm40$ & $0.645\pm0.018$ & $3.385\pm0.221$ & 8.2 & 15 & -0.50 & 1 & \\
J160833.28+311230.8 & 1322022514605143680 & H & DAH & $6420\pm21$ & $0.689\pm0.006$ & $2.708\pm0.065$ & 23.0 & 60 & 0.35 & 1 & \\
J161043.52+132704.9 & 4458251064854338176 & H & DAH & $6600\pm18$ & $0.596\pm0.005$ & $1.851\pm0.021$ & 2.9 & 75 & 0.06 & 1 & \\
J161249.16+325536.0 & 1322796261553165824 & H & DAH & $11482\pm141$ & $0.675\pm0.011$ & $0.518\pm0.19$ & 11.0 & 30 & 0.48 & 1 & \\
J162157.77+043218.8 & 4436905352274528896 & H & DAH & $21636\pm197$ & $1.278\pm0.003$ & $0.690\pm0.018$ & 15 & NA & NA & 4 & \\
J162608.75+055239.9 & 4438653541403499776 & H & DAH & $8912\pm80$ & $0.742\pm0.014$ & $1.196\pm0.041$ & 8.0 & 90 & -0.37 & 1 & X\\
J162808.39+233254.2 & 1299122058220640512 & H & DAH & $7269\pm61$ & $0.782\pm0.014$ & $2.568\pm0.128$ & 4.8 & 19 & -0.21 & 2 & \\
J163036.80+272451.5 & 1304733106575117056 & H & DAH & $11099\pm151$ & $1.054\pm0.013$ & $1.771\pm0.083$ & 35.0 & 45 & 0.30 & 9 & \\
J163917.46+103605.0 & 4447039585308297088 & H & DAH & $6484\pm46$ & $0.787\pm0.017$ & $3.757\pm0.143$ & 1.5 & 17 & -0.21 & 2 & \\
J164358.80+443855.5 & 1405848383457312512 & H & DAH & $5014\pm32$ & $0.572\pm0.020$ & $5.675\pm0.387$ & 3.0 & 45 & -0.30 & 1 & X\\
J165101.72+424955.5 & 1356633384004567168 & log(Ca/He) = -8.03, & DZH & $5569\pm25$ & $0.707\pm0.019$ & $5.599\pm0.159$ & 3.12 & NA & NA & 8 & \\
 & &  log(He/H) = 30.0 & & & & & & & & & \\
J165200.66+411030.0 & 1353302001211658368 & H & DAH & $11073\pm77$ & $0.892\pm0.008$ & $0.973\pm0.022$ & 50.0 & 75 & 0.36 & 1 & \\
J165948.41+440104.5 & 1358301480583401728 & H & DAH & $28170\pm327$ & $1.272\pm0.003$ & $0.292\pm0.012$ & 3.8 & 60 & 0.30 & 9 & \\
J170751.92+353239.4 & 1338455643596995072 & H & DAH & $22332\pm413$ & $1.144\pm0.008$ & $0.300\pm0.017$ & 2.8 & 60 & 0.30 & 9 & \\
J171236.24+392110.1 & 1341599143042727680 & H & DAH & $6043\pm29$ & $0.645\pm0.013$ & $2.783\pm0.109$ & 8.0 & 0 & -0.48 & 1 & \\
J174826.81+005210.0 & 4372558083524803072 & H & DAH & $9065\pm53$ & $0.889\pm0.008$ & $1.841\pm0.060$ & 4.0 & 75 & 0.50 & 1 & \\
J175438.38+384644.6 & 4610983121260366336 & H & DAH & $6431\pm29$ & $0.697\pm0.007$ & $2.789\pm0.078$ & 5.0 & 75 & -0.17 & 1 & \\
J180147.08+652136.7 & 2161573442833511040 & H & DAH & $5425\pm31$ & $0.642\pm0.014$ & $4.482\pm0.246$ & 5.0 & 0 & -0.15 & 1 & \\
J180603.12+464003.2 & 2116551740212362624 & H & DAH & $5816\pm19$ & $0.644\pm0.007$ & $3.215\pm0.079$ & 0.9 & 15 & 0.05 & 1 & X\\
J184926.23+645811.1 & 2253826832091026560 & log(H/He) = -3.29 & DQH? & $12059\pm59$ & $1.048\pm0.006$ & $1.397\pm0.025$ & NA & NA & NA & 9 & \\
J185218.63+183355.4 & 4517521407404432512 & H & DAH & $8759\pm63$ & $0.926\pm0.009$ & $2.362\pm0.070$ & 5.0 & 45 & 0.49 & 1 & X\\
J195629.08-010236.0 & 4235280071072332672 & H & DAH & $7714\pm107$ & $0.689\pm0.022$ & $1.545\pm0.082$ & 0.07 & NA & NA & 11 & \\
J201222.28+311349.5 & 2053953008490747392 & log(H/He) = -6.00 & DBH & $23226\pm505$ & $1.223\pm0.005$ & $0.398\pm0.030$ & 300 & NA & NA & 4 & \\
J201438.48+111157.0 & 1801892849742096128 & H & DAH & $5401\pm26$ & $0.557\pm0.011$ & $3.046\pm0.131$ & 3.5 & 75 & -0.12 & 1 & X\\
J203305.00-183127.2 & 6860084802130167936 & H & DAH & $6225\pm49$ & $0.727\pm0.025$ & $3.553\pm0.243$ & 2.5 & 0 & -0.41 & 1 & X\\
J203524.31-183515.9 & 6861525956933302912 & H & DAH & $12164\pm119$ & $1.138\pm0.007$ & $1.770\pm0.045$ & 104.0 & 15 & -0.10 & 9 & \\
J204400.72+403006.0 & 2066035777984385792 & H & DAH & $7393\pm47$ & $0.850\pm0.010$ & $3.087\pm0.084$ & NA & NA & NA & 3 & \\
J204626.15-071036.9 & 6907031749613795968 & H & DAH & $8603\pm108$ & $0.803\pm0.018$ & $1.543\pm0.092$ & 2.4 & 64 & 0.30 & 7 & \\
J205129.58+035800.8 & 1734159875452345472 & H & DAH & $6993\pm32$ & $0.662\pm0.010$ & $1.869\pm0.037$ & 5.0 & 90 & 0.40 & 1 & \\
J210515.02+090057.4 & 1744102076201595904 & log(Ca/He) = -8.99,  & DZH & $4427\pm31$ & $0.408\pm0.032$ & $4.802\pm0.410$ & NA & NA & NA & 1 & \\
 & & log(He/H) = -30.0 & & & & & & & & & \\
J211125.84+110219.7 & 1745011677261492608 & H & DAH & $16212\pm203$ & $1.128\pm0.007$ & $0.729\pm0.039$ & 3.6 & 60 & 0.30 & 9 & \\
J211504.84+040009.5 & 1732272185785761408 & H & DAH & $6560\pm47$ & $0.664\pm0.022$ & $2.268\pm0.143$ & 4.5 & 6 & -0.23 & 2 & \\
J212232.19-091609.3 & 6897289256251743744 & H & DAH & $6213\pm38$ & $0.773\pm0.014$ & $4.090\pm0.128$ & 9.4 & 15 & -0.35 & 1 & \\
J213148.70+065930.1 & 1739109601921701120 & H & DAH & $6654\pm53$ & $0.738\pm0.026$ & $2.934\pm0.238$ & 31.1 & 9 & -0.25 & 2 & \\
J215135.01+003140.1 & 2681243457490130304 & H & DAH & $8799\pm56$ & $0.908\pm0.013$ & $2.197\pm0.084$ & 300 & NA & NA & 13 & \\
J215843.44+052741.5 & 2697238362376404224 & H & DAH & $9824\pm74$ & $0.875\pm0.012$ & $1.308\pm0.052$ & 3.1 & 60 & 0.26 & 1 & \\
J221141.81+113604.4 & 2727596187657230592 & H & DAH & $9075\pm163$ & $1.312\pm0.008$ & $2.568\pm0.106$ & 18.8 & 22 & -0.35 & 2 & \\
J222109.87-011122.4 & 2677613351131443328 & H & DAH & $5565\pm48$ & $0.717\pm0.032$ & $5.064\pm0.397$ & 9.4 & 0 & -0.34 & 1 & \\
J222348.45+231909.3 & 1878189370339859328 & H & DAH & $8724\pm86$ & $0.916\pm0.014$ & $2.312\pm0.105$ & 6.0 & 15 & -0.50 & 2 & \\
J222740.49+175320.6 & 2737921155893258496 & H & DAZH & $6668\pm30$ & $0.570\pm0.007$ & $1.689\pm0.028$ & 1.3 & 40 & 0.29 & 7 & \\
J223001.91+384630.8 & 1908964185404897152 & H & DAH & $8522\pm62$ & $0.861\pm0.011$ & $2.051\pm0.086$ & 2.9 & 90 & 0.07 & 1 & \\
J225513.49+071001.0 & 2712093451662656256 & H & DAH & $10974\pm208$ & $1.302\pm0.011$ & $2.189\pm0.085$ & 230.0 & 0 & 0.21 & 9 & \\
J225726.05+075541.7 & 2712240064671438720 & H & DAH & $13407\pm128$ & $1.298\pm0.003$ & $1.716\pm0.023$ & 8.8 & 0 & 0.10 & 9 & \\
J225929.73-082822.6 & 2607380156121387648 & H & DAH & $6117\pm25$ & $0.779\pm0.013$ & $4.337\pm0.103$ & 1.4 & 0 & 0.16 & 1 & X\\
J231845.17+123602.7 & 2811321837744375936 & H & DAH & $11616\pm97$ & $0.859\pm0.007$ & $0.783\pm0.020$ & 45 & 80 & 0.05 & 12 & \\
J233203.85+265846.5 & 2865535629374939520 & H & DAH & $9703\pm28$ & $0.891\pm0.004$ & $1.453\pm0.022$ & 2.6 & 57 & 0.30 & 7 & \\
J235006.49-041920.1 & 2447183820215313024 & H & DAH & $5822\pm29$ & $0.739\pm0.018$ & $4.559\pm0.179$ & 2.7 & 0 & 0.00 & 1 & \\
J235032.38-075157.8 & 2436346208978391936 & H & DAH & $5421\pm56$ & $0.760\pm0.028$ & $6.437\pm0.414$ & 6.8 & 15 & -0.32 & 1 & X\\
J235741.37+274727.1 & 2854432108121022464 & log(Ca/He) = -8.51, & DZH & $5750\pm31$ & $0.556\pm0.018$ & $3.013\pm0.186$ & NA & NA & NA & 1 & \\
 & & log(He/H) = 1.0 & & & & & & & & & \\
\enddata
\tablerefs{\scriptsize (1) This Work (2) \citet{Amorim23} (3) \citet{Andrews15} (4) \citet{Bagnulo24} (5) \citet{Bergeron93} (6) \citet{Ferrario15} (7) \citet{Hardy23a} (8) \citet{Hollands17} (9) \citet{Jewett24} (10) \citet{Kulebi09} (11) \citet{Maxted00} (12) \citet{Putney95} (13) \citet{Schmidt03}}
\end{deluxetable}
%\end{longrotatetable}
\end{document}